\documentstyle[pra,aps]{revtex}
\begin{document}
\draft
\title{Dynamical measure and field theory models \\ free of the 
cosmological constant problem}
\author{E. I. Guendelman\thanks{GUENDEL@BGUmail.BGU.AC.IL} and
        A. B. Kaganovich \thanks{ALEXK@BGUmail.BGU.AC.IL}}
\address{Physics Department, Ben Gurion University of the Negev,
   Beer  Sheva 84105, Israel}
\maketitle
\begin{abstract}
 We study field theory models in the context of a gravitational 
theory without the 
cosmological constant problem (CCP). The theory is based on the
requirement 
that the measure of integration  in the action is not necessarily
$\sqrt{-g}$ 
but it is determined dynamically through additional degrees of freedom,
like four scalar fields $\varphi_{a}$.  We study three possibilities for 
the general structure of the
theory: (A) The total action has the form  $S=\int\Phi Ld^{4}x$ where
the measure $\Phi$ is built from the scalars $\varphi_{a}$ in such a
way that the transformation $L\rightarrow L+const$
does not effect equations of motion. 
Then an
infinite dimensional shifts group of the measure fields (SGMF)
$\varphi_{a}$ by arbitrary functions of the Lagrangian density $L$,\quad
$\varphi_{a}\rightarrow \varphi_{a}+f_{a}(L)$, is recognized as the
symmetry group of the action up to an integral of a total divergence.
(B) The total action has the form $S=S_{1}+S_{2}$, $S_{1}=\int\Phi
L_{1}d^{4}x$, $S_{2}=\int\sqrt{-g} L_{2}d^{4}x$ which is the only 
model different from (A) and invariant under SGMF (but now with
$f_{a}=f_{a}(L_{1})$). Similarly, now only $S_{1}$
satisfies
the requirement that the transformation $L_{1}\rightarrow L_{1}+const$
does not effect equations of motion. Both in the case (A) and in the case
(B) it is assumed that $L, L_{1}, L_{2}$ do not depend on $\varphi_{a}$.
(C) The action includes a term which breaks the SGMF symmetry. 
It is
shown
that in the first order formalism, a constraint appears which allows us to
solve the scalar field related to the dynamical measure degrees of
freedom, in terms of matter fields. The remarkable feature of models
discussed in this paper is that for all cases ((A), (B) and (C)), after
the
change of variables to the conformal Einstein frame, the classical field
equations take exactly the form of General Relativity (GR). Therefore the
models are free from the well known problem of the usual scalar-tensor
theories in what is concerned with the classical GR tests. The only
difference of the field equations in the Einstein frame from the canonical
equations of the selfconsistent system of Einstein's gravity and matter
fields, is the appearance of the effective scalar field potential which
vanishes in a true vacuum   
state (TVS) without fine tuning in cases (A) and (B). To illustrate how 
the
theory works, we present a few explicit
field theory models where it is possible to combine the solution of the
cosmological constant problem
(CCP) with: 1) possibility for inflationary scenario; 2) spontaneously
broken gauge unified theories (including fermions). 
In the case (C),
the
breaking of the SGMF symmetry induces a nonzero energy density for the
TVS.
When considering only
a linear potential for a
scalar field $\phi$ in $S_{1}$, the continuous symmetry
$\phi\rightarrow\phi+const$ is respected. Surprisingly, in this case SSB
takes place while no massless ("Goldstone") boson appears.
We discuss the role of the SGMF symmetry for quantization and 
the possible connection
of this theory with theories of extended objects.

\renewcommand{\baselinestretch}{1.6}
PACS number(s): 11.15.Ex, 98.80.Cq, 12.10.Dm, 04.90.+e
\end{abstract}

 \pagebreak
\section{Introduction}

The cosmological constant problem in the context of general relativity 
(GR) can be explained as follows. In GR one can introduce such a constant 
or one may set it to zero. The problem is that after further 
investigation of elementary particle theory, we discover new phenomena 
like radiative corrections, the existence of condensates, etc. each of 
which contributes to the vacuum energy. In order to have a resulting zero 
or extremely small cosmological constant as required by observations of 
the present day universe , one would have to carefully fine tune 
parameters in the Lagrangian so that all of these contributions more or 
less exactly cancel. This question has captured the attention of many 
authors because, among other things, it could be a serious indication 
that something fundamental has been missed in our standard way of 
thinking about field theory and the way it must couple to gravity. For a 
review of this problem see \cite{CC}.

The situation is made even more serious if one believes in the existence 
of an inflationary phase for the early universe, where the vacuum energy 
plays an essential role. The question is then: what is so special about 
the present vacuum state which was not present in the early universe? In 
this paper we are going to give an answer this question. 

As it is well known, in nongravitational physics the origin from which we
measure energy is not important. For example in nonrelativistic mechanics
a shift in the potential $V\rightarrow V+constant$ does not lead to any
consequence in the equations of motion. In the GR the situation changes 
dramatically. There {\em all} the
energy density, including the origin from which we measure it, affects the
gravitational dynamics.

This is quite apparent when GR is formulated from a variational approach.
There the action is
\begin{equation}
S=\int\sqrt{-g}Ld^{4}x
\label{I1}
\end{equation}
\begin{equation}
L=-\frac{1}{\kappa}R(g)+L_{m}
\label{I2}
\end{equation}
where $\kappa=16\pi G$, $R(g)$ is the Riemannian scalar curvature of the
4-dimensional space-time with metric $g_{\mu\nu}$, $g\equiv
Det(g_{\mu\nu})$ and $L_{m}$ is the matter Lagrangian density.
It is apparent now that the shift of the Lagrangian density $L$,
\quad $L\rightarrow L+C$,\quad $C=const$ is not a symmetry of the action 
(\ref{I1}).
Instead, it leads to an additional piece in the action of the form
$C\int\sqrt{-g}d^{4}x$ which contributes to the equations of motion and
in particular generates a so called "cosmological constant term" in the
equations of the gravitational field. 

In Refs. \cite{GK1}-\cite{GK4} an approach has been 
developed where the
cosmological constant problem is treated as the absence of gravitational
effects of a possible constant part of the Lagrangian density. The basic
idea is that the measure of integration in the action principle is not
necessarily $\sqrt{-g}$ but it is allowed to "float" and to be determined
dynamically through additional degrees of freedom. In other words the
floating measure is not from first principles related to $g_{\mu\nu}$, 
although relevant equations will in general allow to solve for the new 
measure in terms of other fields of the theory ($g_{\mu\nu}$ and matter 
fields). This theory is based on the demand that such measure respects  
{\em the principle of
non gravitating vacuum energy} (NGVE principle) which states that the
Lagrangian density $L$ can be changed to $L+constant$ without affecting
the dynamics. This requirement is imposed in order to offer a new
approach for the solution of the cosmological constant problem.
Concerning the theories based on the NGVE principle we will refer to
them  as NGVE - theories. 

The invariance $L\longrightarrow L+constant$ for the action is
achieved if the measure of integration in the action is a total
derivative, so that
to an infinitesimal hypercube in 4-dimensional space-time $x_{0}^{\mu}\leq
x^{\mu}\leq x_{0}^{\mu}+dx^{\mu}$, $\mu =0,1,2,3$ we associate a volume
element $dV$ which is: (i) an exact differential, (ii) it is proportional
to
$d^{4}x$ and (iii) $dV$ is a general coordinate invariant. The usual choice,
$\sqrt{-g} d^{4}x$ does not satisfy condition (i).

The conditions (i)-(iii) are satisfied \cite{GK1}, \cite{GK2} if
the measure corresponds to the integration in the space of the four 
scalar fields  $\varphi_{a}, (a=1,2,3,4)$, that is
\begin{equation}
dV =
d\varphi_{1}\wedge
d\varphi_{2}\wedge
d\varphi_{3}\wedge
d\varphi_{4}\equiv\frac{\Phi}{4!}d^{4}x
\label{dV}
\end{equation}
where
\begin{equation}
\Phi \equiv \varepsilon_{a_{1}a_{2}a_{3}a_{4}}
\varepsilon^{\mu\nu\lambda\sigma}
(\partial_{\mu}\varphi_{a_{1}})
(\partial_{\nu}\varphi_{a_{2}})
(\partial_{\lambda}\varphi_{a_{3}})
(\partial_{\sigma}\varphi_{a_{4}}).
\label{Fi}
\end{equation}
is a measure independent of $g_{\mu\nu}$ as opposed to the case of GR. In
what follows we will call $\Phi$ the {\em dynamical measure} since its
value in realistic cases is dynamically determined in terms of the other
fields of the theory through the equations of motion as we will see. We
will use also the terms the "dynamical" volume element or the volume
element of the "dynamical space - time" for the expression $\Phi d^{4}x$
since according to Eq. (\ref{dV}), it is actually the volume element of
the internal space of fields $\varphi_{a}$ and in the action it will take
the place of the space - time volume element of GR.

Notice that the dynamical  measure (\ref{Fi}) is a particular
realization of the NGVE-principle 
(for other possible realization see Refs. \cite{GK3}, \cite{GK4}).
For additional discussion of the geometrical meaning of this realization
of the 
measure see Ref.\cite{Hehl} 

We will study three possibilities for the general structure of the action:

(A) The most straightforward and complete realization of the NGVE
principle (the so called strong NGVE principle) is a theory where the
total action is just
defined as follows
\begin{equation}
S =\int L\Phi d^{4}x
        \label{Action}
\end{equation}
where $L$ is a total Lagrangian density. We assume in what follows that 
$L$ does not contain explicitly the measure fields, that is the fields
$\varphi_{a}$ by means of which $\Phi$ is defined.

Introducing independent degrees of freedom related to the dynamical 
measure we 
arrive naturally at a conception that degrees of freedom associated with 
all possible geometrical objects
(like metric, connection and measure) that 
 appear, should be considered as independent ones. This is why we expect
that the 
first order formalism, where the affine connection is {\em not} assumed to 
be the Christoffel coefficients in general should be preferable to the 
second order formalism where this assumption is made. 

In fact, it is found \cite{GK4}, \cite{GK5},
\cite{PASCOS}, \cite{Fried98} that the NGVE theory in the context of the
first order 
formalism does indeed provide: \\
1) A possibility to construct models with
the same number of degrees of freedom as in GR and with the same structure
of equations after a suitable change of variables. In particular, it means
that in the contrast to the usual scalar - tensor theories, a  conformal
frame  exists where both the gravitational constant and all masses are
constant simultaneously. It implies also that
gravitational radiation for example will be the same as that in GR. By the
same reason, in the true vacuum, the only spherically symmetrical solution
is Schwarzschild. Therefore, the classical GR tests are guaranteed to work
out correctly. \\
2) A solution of the cosmological constant problem. \\
We have to emphasize
that this is not the case when using the second 
order formalism \cite{GK1}, \cite{GK2}. 

The simplest example \cite{GK3}, \cite{GK4} (see also Sec.IIC)
where these 
ideas can be tested is that of a matter Lagrangian described by a single 
scalar field with a nontrivial potential. In this case the variational 
principle leads to a constraint which implies the vanishing of the 
effective vacuum energy in any possible allowed configuration of the 
scalar field. These allowed configurations are however constant values at 
the extrema of the scalar field potential and an integration constant 
that results from the equations of motion has the effect of exactly 
canceling the value of the potential at these points. So, the scalar 
field is forced to be a constant and hence  
the theory has no nontrivial dynamics for the scalar field.

In this case the measure (\ref{Fi}) is not determined by the equations of 
motion. In fact a local symmetry (called "Local Einstein Symmetry" (LES)) 
exists which allows us to choose the measure $\Phi$ to be of
whatever we 
want. In 
particular $\Phi=\sqrt{-g}$ can be chosen and in this case the theory 
coincides in the vacuum with GR with $\Lambda =0$.

A richer structure is obtained if a four index field strength which 
derives from a three index potential is allowed in the theory as it is 
demonstrated in Appendixes C and D and in Sec.IID (see also Refs. 
\cite {GK4}, \cite{GK5} - \cite{Fried98}). The 
introduction of this term breaks the LES mentioned above. In this case, 
the constraint that the theory provides, allows  to solve for the measure
$\Phi$
in terms of $\sqrt{-g}$ {\em and} the matter fields of the theory. 
It turns out that by the use of a conformal transformation the 
equations can be written in the form  of the Einstein 
theory equations where the resulting particle interactions and potentials
are replaced by effective ones. Since the  picture in these new variables
has the structure of the Einstein's GR we will call this conformal frame
the Einstein frame and we will refer to this picture as "Einstein
picture".

Then the theory which contains a scalar field and  a four index field 
strength shows a remarkable
feature: 
the effective potential of the scalar field that one obtains in the 
Einstein  picture 
is such that  generally allows for an inflationary phase which
evolves at 
a later stage, 
without fine tuning, to a  vacuum of the theory  with zero cosmological 
constant \cite {GK4}, \cite{GK5} - \cite{Fried98}. 

It is interesting to notice the existence of the symmetry of the 
classical equations of motion under the global transformation 
$\Phi\rightarrow\alpha\Phi$, $\alpha =constant$ (which can be obtained by
individual global 
rescalings of the $\varphi_{a}$'s like 
$\varphi_{a}\rightarrow\lambda^{1/4}\varphi_{a}$). In the Einstein frame 
this symmetry includes a global scale transformation of the metric. The 
expectation value of 4-index field strength spontaneously breaks this 
scaling symmetry without generating a massless dilaton.

The 4-index field strength also allows  for a Maxwell-type dynamics
of 
gauge fields \cite{PASCOS}, \cite{Fried98} and of massive fermions.
 For illustration how the
theory works, the model which provides a possibility for  an explicit
construction
of unified gauge theory ($SU(2)\times U(1)$ as an example) based on 
these ideas and keeping all the above mentioned advantages, is 
presented in Sec.V.         

(B) It is possible to check (see \cite{GK1} and Sec. IIA) that the action 
(\ref{Action})
respects (up to an integral of a total divergence) the
infinite dimensional group of shifts of the measure fields $\varphi_{a}$
(SGMF)
\begin{equation}
\varphi_{a}\rightarrow \varphi_{a}+f_{a}(L)
        \label{SG}
\end{equation} 
where $f_{a}(L)$ is an {\em arbitrary} differentiable function of the
total Lagrangian density $L$. Such symmetry in general represents a
nontrivial mixing between the measure fields $\varphi_{a}$ and the matter
and gravitational fields (through L). As it was mentioned in Ref.
\cite{GK1}, this symmetry prevents the appearance of terms of the form
$f(\chi)\Phi$
in the effective action (where quantum corrections are taken into account)
with the single possible exception of $f(\chi)=
c/\chi$ where a scalar $c$ is $\chi$ independent. This
is because in this last  case
the term $f(\chi)\Phi =c\sqrt{-g}$ is $\varphi_{a}$ independent. This
possibility gives rise to the cosmological constant term  in the action 
 while the symmetry (6) is maintained. This
can be generalized to possible contributions of the form
$\int L_{2}\sqrt{-g}d^{4}x$ where $L_{2}$ is $\varphi_{a}$ independent
function of matter fields and gravity if radiative corrections generate a
term $f(\chi)\Phi$ with $f(\chi)=L_{2}/\chi$.

So, let us consider an action which
consists of two terms 
\begin{eqnarray}
S & = & S_{1}+S_{2}
\nonumber\\
S_{1} & = & \int L_{1}\Phi d^{4}x
\nonumber\\
S_{2} & = & \int L_{2}\sqrt{-g}d^{4}x
        \label{Action12}
\end{eqnarray}

Now only $S_{1}$ satisfies the requirement that the transformation
$L_{1}\rightarrow L_{1}+const$ does not effect equations of motion, which
is a somewhat weaker version of the NGVE principle. 
In this case, the SGMF
symmetry transformation (\ref{SG}) is replaced by 
\begin{equation}
\varphi_{a}\rightarrow \varphi_{a}+f_{a}(L_{1})
        \label{SG1}
\end{equation}

The constraint which appears again in the first order formalism, allows
now to solve
the measure
$\Phi$
in terms of $\sqrt{-g}$ {\em and} the matter fields of the theory
without introducing the four index field strength (see Sec. III). 

In scalar field models with potentials entering in $S_{1}$ and $S_{2}$,
in
the true vacuum state (TVS) $\chi\rightarrow\infty$. However, in the
Einstein frame this singularity does not present and the energy
density of TVS is zero without fine tuning of any  scalar potential in
$S_{1}$ or $S_{2}$ (see Ref.\cite{Fried98}). This means that even the weak
version of the NGVE
principle is enough to provide a solution of the CCP. 
We show  in
Subsec. IIIB how it is possible to incorporate gauge fields in such kind
of
model.

Sections IV and V are devoted to the ways for inclusion of fermions and
construction of unified gauge theories in the context of the NGVE
theory.

(C) As it is well known, in order to really understand the role of some
symmetry, one should see what the breaking of such symmetry does. 
With a simple example in Sec.VI, we will see that the breaking of the
symmetry
(\ref{SG}) or (\ref{SG1}) can lead to the appearance of a nonzero energy
density for the
TVS. In the particular example we study, the additional piece in the
action that is added is of the form
\begin{equation}
S_{3}=-\gamma\int\frac{\Phi^{2}}{\sqrt{-g}}d^{4}x.
        \label{Break}
\end{equation}
which is equivalent to considering of a piece of the Lagrangian
density $L_{1}$ linear in $\chi$. Then
as we show in Sec.VI, the TVS energy density appears to be equal to 
$\gamma$.

Notice that for the field theory models presented  in this paper in the
framework of  the case (B)
as well as of the case (C), the classical GR tests are guaranteed
to work
out correctly by the same resons as in the case (A) (provided that in the
case (C), $\gamma$
in Eq. (\ref{Break}) is small enough).

In Sec.VII we show that when considering only
a linear potential for a
scalar field $\phi$ in $S_{1}$, the continuous symmetry
$\phi\rightarrow\phi+const$ is respected. Surprisingly, in this case SSB
takes place while no massless ("Goldstone") boson appears. Models with
such a feature exist in each of the cases (A), (B), (C).

\bigskip
\section{Models satisfying the strong NGVE principle}
\subsection{General features of the strong NGVE theory}

\bigskip

Starting from the case (A) (see Introduction), we consider the action 
(\ref{Action}). We assume that the total Lagrangian density $L$ in Eq. 
(\ref{Action}) does not contain the measure fields $\varphi_{a}$, that 
is the fields by means of which the measure $\Phi$ is defined. If this 
condition is
satisfied then the action (\ref{Action}) is 
invariant, up to a total divergence, under transformations (\ref{SG})
(see also \cite{GK1}). In fact, by the transformation (\ref{SG}) the action 
(\ref{Action}) is changed according to $S\rightarrow S+\delta S$ with
\begin{equation}
\delta S=4\int A^{\mu}_{a}L\partial_{\mu}f_{a}(L)d^{4}x+
6\int\varepsilon^{\mu\nu\alpha\beta}\varepsilon_{abcd}L
\partial_{\mu}\varphi_{a}
\partial_{\nu}\varphi_{b}
\partial_{\alpha}L
\partial_{\beta}L\frac{df_{c}}{dL}\frac{df_{d}}{dL}
+\ldots 
 \label{SG2}  
\end{equation}
where
\begin{equation}
A^{\mu}_{a}=\varepsilon_{bcda}
\varepsilon^{\nu\lambda\sigma\mu}
(\partial_{\nu}\varphi_{b})
(\partial_{\lambda}\varphi_{c})
(\partial_{\sigma}\varphi_{d})
\label{DefA}
\end{equation}
and all terms in the right hand side of Eq.(\ref{SG2}), with the exception
of the first one, are 
identically equal to zero because of symmetry properties. The first term 
is transformed to the total divergence
\begin{equation}
\delta S=\int\partial_{\mu}\Omega^{\mu}d^{4}x 
\label{SG3}
\end{equation}
where $\Omega^{\mu}\equiv 4A^{\mu}_{a}g_{a}(L)$ and $g_{a}(L)$ being 
defined from $f_{a}(L)$ through the equation $L(df_{a}/dL)=dg_{a}/dL$ 
and the identity $\partial_{\mu}A^{\mu}_{a}\equiv 0$ has been taken into 
account.

Our choice for the total Lagrangian
density is
\begin{equation}
L=-\frac{1}{\kappa}R(\Gamma,g)+L_{m}
\label{L1}
\end{equation}
where $L_{m}$ is the matter Lagrangian density and $R(\Gamma,G)$ is the
scalar  curvature which in the first order formalism in the framework of 
the Metric-Affine theory \cite{HN} is defined as follows
 \begin{equation}
R(\Gamma,g)=g^{\mu\nu}R_{\mu\nu}(\Gamma)
\label{R}
\end{equation}

\begin{equation}
R_{\mu\nu}(\Gamma)=R^{\lambda}_{\mu\nu\lambda}(\Gamma)
\label{RAB}
\end{equation}

\begin{equation}
R^{\lambda}_{\mu\nu\sigma}(\Gamma)\equiv 
\Gamma^{\lambda}_{\mu\nu ,\sigma}-\Gamma^{\lambda}_{\mu\sigma ,\nu}+
\Gamma^{\lambda}_{\alpha\sigma}\Gamma^{\alpha}_{\mu\nu}-
\Gamma^{\lambda}_{\alpha\nu}\Gamma^{\alpha}_{\mu\sigma}
\label{RABCD}
\end{equation}  
where $\Gamma^{\lambda}_{\mu\nu}$ are the connection coefficients which 
have to be obtained from the variational principle.

Equations that originate from the variation of the action (\ref{Action}) 
with respect to the measure fields $\varphi_{a}$, are  
\begin{equation}
A^{\mu}_{a}\partial_{\mu}\lbrack -\frac{1}{\kappa}R(\Gamma,g)+
L_{m}\rbrack =0
\label{FEM}
\end{equation}

Since
$A_{a}^{\mu}\partial_{\mu}\varphi_{a^{\prime}}=4^{-1}\delta_{aa^{\prime}}\Phi$
we get \footnote{For the connection with other developments
concerning volume preserving diffeomorphisms and the corresponding
structure of fields equations see Appendix A} that $Det (A^{\mu}_{a})
= \frac{4^{-4}}{4!}\Phi^{3}$, so that if $\Phi\neq 0$, it follows from 
Eq.(\ref{FEM})
\begin{equation}
 -\frac{1}{\kappa}R(\Gamma,g)+
L_{m}=M=constant
\label{II1}
\end{equation}

Let us now study equations that originate from variation with respect to 
$g^{\mu\nu}$. For simplicity we present here the calculations for the 
case where there are no fermions. Performing the variation with respect 
to $g^{\mu\nu}$ we get
\begin{equation}
 -\frac{1}{\kappa}R_{\mu\nu}(\Gamma)+
\frac{\partial L_{m}}{\partial g^{\mu\nu}}=0
\label{II2}
\end{equation}

Contracting Eq.(\ref{II2}) with $g^{\mu\nu}$ and making use 
Eq.(\ref{II1}) we get
\begin{equation}
g^{\mu\nu}\frac{\partial (L_{m}-M)}{\partial g^{\mu\nu}}-
(L_{m}-M)
=0
\label{II3}
\end{equation}

This equation is a constraint since generically $L_{m}$ contains only the 
fields and their first derivatives. A similar constraint is achieved by 
using the vierbein - spin-connection (VSC) formalism (see Appendix E). 
Notice that if $L_{m}-M$ is homogeneous of degree one in $g^{\mu\nu}$ it 
satisfies the constraint (\ref{II3}) automatically (that is without using 
equations of motion).

\bigskip
\subsection{The vacuum case}
In the vacuum, when we choose $L_{m}=0$ in Eq.(\ref{L1}), it follows from 
Eq.(\ref{II2})
\begin{equation}
R_{\mu\nu}(\Gamma)
=0
\label{III1}
\end{equation}

Eq.(\ref{II1}) implies then that the integration constant $M=0$. Adding 
an arbitrary constant $C$ to $L_{m}$ does not change the resulting 
Eq.(\ref{III1}) since, as we see from Eqs. (\ref{II1}) and (\ref{II2}) or 
from the constraint (\ref{II3}), the integration constant $M$ has to 
compensate the constant $C$.

To clarify the sense of Eq.(\ref{III1}) we need the connection 
coefficients $\Gamma^{\lambda}_{\mu\nu}$. Varying the action (\ref{Action})
with $L=-\frac{1}{\kappa}R(\Gamma ,g)$ with respect to 
$\Gamma^{\lambda}_{\mu\nu}$,  we get
\begin{eqnarray}
-\Gamma^{\lambda}_{\mu\nu}-
\Gamma^{\alpha}_{\beta\mu}g^{\beta\lambda}g_{\alpha\nu}+
\delta^{\lambda}_{\nu}\Gamma^{\alpha}_{\mu\alpha}
+\delta^{\lambda}_{\mu}g^{\alpha\beta}\Gamma^{\gamma}_{\alpha\beta}
g_{\gamma\nu}-
\nonumber\\
g_{\alpha\nu}\partial_{\mu}g^{\alpha\lambda}+
\delta^{\lambda}_{\mu}g_{\alpha\nu}\partial_{\beta}g^{\alpha\beta}-
\delta^{\lambda}_{\nu}\frac{\Phi,_{\mu}}{\Phi}+
\delta^{\lambda}_{\mu}\frac{\Phi,_{\nu}}{\Phi}
=0.
\label{GAM1}
\end{eqnarray}

We will look for the solution 
of the form
\begin{equation}
\Gamma^{\lambda}_{\mu\nu}=\{ ^{\lambda}_{\mu\nu}\}+\Sigma^{\lambda}_{\mu\nu}
\label{GAM2}
\end{equation}
where $\{ ^{\lambda}_{\mu\nu}\}$  are the Christoffel's connection 
coefficients. Then $\Sigma^{\lambda}_{\mu\nu}$ satisfies the equation
\begin{equation}
-\sigma,_{\lambda}g_{\mu\nu}+\sigma,_{\mu}g_{\nu\lambda}-
g_{\nu\alpha}\Sigma^{\alpha}_{\lambda\mu}-
g_{\mu\alpha}\Sigma^{\alpha}_{\nu\lambda}+
g_{\mu\nu}\Sigma^{\alpha}_{\lambda\alpha}+
g_{\nu\lambda}g_{\alpha\mu}g^{\beta\gamma}\Sigma^{\alpha}_{\beta\gamma}=0
\label{S1}
\end{equation}
where
\begin{equation}
\sigma\equiv\ln\chi, \hspace{1.5cm} \chi\equiv\frac{\Phi}{\sqrt{-g}}
\label{ski}
\end{equation}

The general solution of Eq.
(\ref{S1}) is
\begin{equation}
\Sigma^{\alpha}_{\mu\nu}=\delta^{\alpha}_{\mu}\lambda,_{\nu}+
\frac{1}{2}(\sigma,_{\mu}\delta^{\alpha}_{\nu}-
\sigma,_{\beta}g_{\mu\nu}g^{\alpha\beta})
\label{S2}
\end{equation}
where $\lambda$ is an arbitrary function which appears due to the 
existence of the Einstein - Kaufman $\lambda$-symmetry (see \cite{EK}, 
\cite{GK3} and Appendix B): the curvature tensor (\ref{RABCD}) is 
invariant under the $\lambda$- transformation
\begin{equation}
\Gamma^{\prime
\alpha}_{\mu\nu}(\lambda, 
\sigma)=\Gamma^{\alpha}_{\mu\nu}+\delta^{\alpha}_{\mu}\lambda,_{\nu} 
\label{Gamal} 
\end{equation}
Although this symmetry was discussed in Ref.\cite{EK} in a very specific 
unified theory, it turns out that $\lambda$- symmetry has a wider range 
of validity and in particular it is useful in our case. 

If we choose the gauge $\lambda=\frac{\sigma}{2}$, then the antisymmetric 
part of $\Sigma^{\alpha}_{\mu\nu}$ disappears and we get 
\begin{equation}
\Sigma^{\alpha}_{\mu\nu}(\sigma)=\frac{1}{2}(\delta^{\alpha}_{\mu}\sigma,_{\nu}
+\delta^{\alpha}_{\nu}\sigma,_{\mu}-
\sigma,_{\beta}g_{\mu\nu}g^{\alpha\beta})
\label{S3}
\end{equation}
which contributes to the nonmetricity (more discussion about $\lambda$
symmetry and other gauge choices see Appendix B).

In the vacuum, the $\sigma$-contribution (\ref{S3}) to the
nonmetricity  can be eliminated. This is because in the
vacuum ($L_{m}=0$) the action (\ref{Action}), (\ref{L1}) is invariant
under the local 
Einstein symmetry (LES)
 \begin{equation}
g_{\mu\nu}(x)=J^{-1}(x)g^{\prime}_{\mu\nu}(x)
\label{ES1}
\end{equation}
\begin{equation}
\Phi(x)=J^{-1}(x)\Phi^{\prime}(x)
\label{ES2}
\end{equation}

The transformation (\ref{ES2})
can be the result of a diffeomorphism
$\varphi_{a}\longrightarrow\varphi^{\prime}_{a}=
\varphi^{\prime}_{a}(\varphi_{b})$ in the space of the scalar fields
$\varphi_{a}$ (see Ref. \cite{GK2}). Then $J=
Det(\frac{\partial\varphi^{\prime}_{a}}{\partial\varphi_{b}})$.

Notice that even when we are not in the vacuum, but the matter Lagrangian 
density $L_{m}$ satisfies the constraint (\ref{II3}) automatically (that 
is $L_{m}$ is homogeneous of degree one in $g^{\mu\nu}$, up to irrelevant 
additive constant) then the total action  (\ref{Action}), (\ref{L1}) 
possesses LES too. For examples see \cite{GK2},\cite{GK3}.  

 For $J=\chi$ we get $\chi^{\prime}\equiv 1$,
\quad $\Sigma^{\prime \alpha}_{\mu\nu}(\sigma)\equiv 0$ and
$\Gamma^{\prime 
\alpha}_{\mu\nu}= \{ ^{\alpha}_{\mu\nu}\}^{\prime}$, where
$\{ ^{\alpha}_{\mu\nu}\}^{\prime}$
 are the Christoffel's coefficients corresponding
to the new metric $g^{\prime}_{\mu\nu}$.
In terms of the new metric $g^{\prime}_{\mu\nu}$, the curvature 
(\ref{RABCD}) becomes the Riemannian curvature and therefore Eq. (\ref{III1})
is equivalent to the vacuum Einstein's equation with zero cosmological 
constant.

\bigskip
\subsection{Single scalar field with a nontrivial potential}

Now let us consider the cases when the constraint (\ref{II3}) is not satisfied
without restrictions on the dynamics of the matter fields. Nevertheless,
the constraint (\ref{II3}) holds as a consequence of the variational
principle in any situation.

A simple case where the constraint (\ref{II3}) is not automatic is the case
of a single scalar field with a nontrivial potential $V(\varphi)$
\begin{equation}
L_{m}= \frac{1}{2}\varphi_{,\alpha}\varphi^{,\alpha}-V(\varphi)
 \label{III2}
 \end{equation}
In this model, the kinetic part of the action possesses LES and satisfies 
the constraint automatically since 
$\frac{1}{2}\varphi_{,\alpha}\varphi^{,\alpha}$ is homogeneous of degree 
one in 
$g^{\mu\nu}$. The potential part apparently does not satisfy the LES and 
as a result of this the constraint (\ref{II3}) implies
\begin{equation}
V(\varphi)+M=0
 \label{CS}
 \end{equation}
Therefore we conclude that, provided $\Phi\neq 0$, there is no dynamics for
the theory of a single scalar field with a nontrivial potential, since
constraint (\ref{CS}) forces
this scalar field to be a constant. 

 The constraint (\ref{CS}) has to be
solved together with the equation of motion resulting from the variation
of the action (\ref{Action}), (\ref{L1}), (\ref{III2}) with respect to
$\varphi_{a}$
\begin{equation}
(-g)^{-1/2}\partial_{\mu}(\sqrt{-g}g^{\mu\nu}\partial_{\nu}\varphi)+
\sigma,_{\mu}\varphi ^{,\mu}+\frac{dV}{d\varphi}=0,
 \label{SE}
 \end{equation}
where $\sigma =\ln\chi$. From eqs.(\ref{CS}) and (\ref{SE}) we conclude
that the $\varphi$ -field has to be
located at an extremum of the potential
$V(\varphi)$. Since the constraint(\ref{CS}) eliminates the dynamics of the
scalar field $\varphi$, we cannot really say that we have a situation where
the LES (\ref{ES1}), (\ref{ES2}) is actually broken, since after solving
the constraint together with the equation of motion (i.e. on the mass 
shell) the symmetry remains true.

Taking into account that $\varphi =constant$ and 
$L_{m}-M=-(V(\varphi)+M)=0$, we 
see from Eqs. (\ref{II1})-(\ref{II3}) that $R_{\mu\nu}(\Gamma,g)=0$. As 
we have seen in Sec. IIB, the $\sigma$ contribution to the 
connection can be eliminated in the vacuum by the transformations 
(\ref{ES1}), (\ref{ES2}). Notice that since the equations of motion
enforce $\varphi =constant$, the 
single scalar field $\phi$ part of the Lagrangian density acts as 
an arbitrary constant. However, this 
situation is indistinguishable from the vacuum case. Then repeating the 
LES transformation of the end of Sec. IIB, we see that in 
terms of the new metric $g^{\prime}_{\mu\nu}$, 
the tensor $R_{\mu\nu}(\Gamma,g)$ becomes the usual Ricci tensor 
$R_{\mu\nu}(g^{\prime})$ of the Riemannian space-time with the metric 
$g^{\prime}_{\mu\nu}$ and $R_{\mu\nu}(g^{\prime})\equiv 0$
Therefore we conclude that {\em for the case of a single scalar field with a 
nontrivial potential, the theory is equivalent to the Einstein's GR with 
the zero cosmological constant}. As a consequence, in this simple model, 
among maximally symmetric solutions, only Minkowski space is a solution. 
The absence of de Sitter space as a solution makes us suspect that the 
NGVE theory is inconsistent with the idea of inflation. This however is 
not true as we will see in the next subsection.

\bigskip
\subsection{Four index field strength, unified gauge sector and a model with 
realistic particle fields dynamics and cosmology}

\bigskip

As we have mentioned, one of the biggest puzzles of modern physics is
what is referred to as the
"cosmological constant problem", i.e. the absence of a possible constant
part of the vacuum energy in the present day universe \cite{CC}. On the
other hand, many questions in modern
cosmology appear to be solved by the so called "inflationary model" which
makes use of a big effective cosmological constant in the
early universe \cite{Gut}. A possible conflict between a successful
resolution of the cosmological constant problem and the existence of
an inflationary phase could be a "potential Achilles heel for the
scenario" as has been pointed out\cite{KT}. Here we will show (see also
\cite{GK4}) that indeed there is no conflict between the existence of an
inflationary phase and the disappearance of the cosmological constant  in
the later phases of cosmological evolution (without the need of fine
tuning). In the context of the strong NGVE theory, the introduction of a
four index field strength condensate plays a 
crucial role for this.

Another problem related to the NGVE theory consists of the very strong
restriction which constraint (\ref{II3}) dictates on the matter models
which generally do not satisfy the LES. This makes the incorporation of
 fermion masses and gauge fields not straightforward (see Refs.
\cite{GK2}, \cite{GK3}). In what follows we will see, however, that the
incorporation of the four index-field strength in four dimensional
space - time turns the constraint into an equation for $\chi$- field.
After
solving this constraint we obtain well defined matter models.

In Appendixes C and D we show how  it is possible to realize a nontrivial 
dynamics of a scalar field and a gauge field while solving the CCP. 
However, the simplest models presented there give rise to undesirable 
problems: either the mass of the scalar field turns out to be infinite 
 or nonminimal nonrenormalizable couplings appear at very high energies.
We 
are going to show in this section that by a certain sort of 
{\em unification of all gauge fields} we can get rid of the above 
mentioned problem. This approach seems to be the most appealing one in 
the context of the theory based on the strong NGVE principle. This 
Section is self contained, but Appendixes C and D help the reader to 
understand how the theory operates generically. 

The basic idea consists of demanding that the dependence of the 
Lagrangian density on the gradients of the gauge field potentials is only 
through a single variable which is the sum of all possible kinetic terms 
and a corresponding four index field strength term. The fact that all 
gauge fields must come together is automatic in a unified gauge models 
where the Lagrangian density must depend only on 
$F_{\mu\nu}^{a}F^{a\mu\nu}$ where $a$ is for example an $SU(5)$ index. In 
addition we insist also in introducing a three index potential 
$A_{\mu\nu\alpha}$ into the game in a similar way. Here we will demand 
that the field strength 
$F_{\alpha\mu\nu\beta}=\partial_{[\alpha}A_{\mu\nu\beta]}$ 
is combined with other usual gauge field kinetic terms, like 
$F_{\mu\nu}^{a}F^{a\mu\nu}$, in such a way that the homogeneity of each 
of the terms in $g^{\mu\nu}$ is of degree 2. This singles out the 
following combination analytic in the gradients of the gauge fields 
potentials 
\footnote {The other possibility including
$\sqrt{-F_{\mu\nu\alpha\beta}F^{\mu\nu\alpha\beta}}\equiv
|\frac{\varepsilon^{\mu\nu\alpha\beta}}{\sqrt{-g}}
\partial_{\mu}A_{\nu\alpha\beta}|$ is not analytic one.}
\begin{equation}
y\equiv F_{\mu\nu}F^{\mu\nu}+m^{2}\frac{\varepsilon^{\mu\nu\alpha\beta}} 
{\sqrt{-g}} \partial_{\mu}A_{\nu\alpha\beta}
 \label{V1}
\end{equation}
We will call $y$ the gauge complex. Here  $m$ is a parameter with 
the dimensions of mass.

The demand that the term which depends on the condensate of 
$A_{\mu\nu\alpha}$, has to have the same transformation under 
$g^{\mu\nu}\rightarrow \Omega g^{\mu\nu}$ as the ordinary gauge fields, 
finds a simple analogy in a related higher dimensional picture where the 
components of the gauge field strengths in the direction of the extra 
dimensions can play a similar role to that of the $A_{\mu\nu\alpha}$ 
field in 4-dimensional space-time. In 6-dimensional case for example, 
$F_{AB}F^{AB}\equiv 
F_{\mu\nu}F^{\mu\nu}+2F_{a\mu}F^{a\mu}+F_{ab}F^{ab}$, where $\mu,\nu 
=0,1,2,3$; \ $a,b$=4,5 and $F_{ab}F^{ab}$ plays then the role of 
$\frac{\varepsilon^{\mu\nu\alpha\beta}}{\sqrt{-g}}
\partial_{\mu}A_{\nu\alpha\beta}$  in the condensate state \cite{G} (here 
$F_{ab}$ takes a "magnetic monopole" expectation value). Then the 
requirement of equal behavior under conformal transformation is of course 
automatic.

To implement the suggestion of using the combination (\ref{V1}) let us 
consider a model with the action 
\begin{eqnarray}
S=\int\Phi d^{4}x[-\frac{1}{\kappa}R(\Gamma,g)-m^{4}f(u)+
\frac{1}{2}g^{\mu\nu}\phi,_{\mu}\phi,_{\nu}
+g^{\mu\nu}(\partial_{\mu}-i\tilde{e}\tilde{A}_{\mu})\phi
           (\partial_{\nu}+i\tilde{e}\tilde{A}_{\nu})\phi^\ast
-V(|\phi|)]
 \label{W18}
\end{eqnarray}
where $\phi$ is a complex scalar field minimally coupled to a vector gauge
field $\tilde{A}_{\mu}$ and $f(u)$ is a
function of the dimensionless argument $u=y/m^{4}$. We will see that the
only requirement condition on the function $f(u)$, that provides a
persistent condensate with physically reasonable consequences, is that
$f^{\prime}(u)\equiv\frac{df}{du}=0$ for some $y=y_{0}>0$.

By making use the gauge invariance we choose the unitary gauge (where
$Im\,\phi (x)=0$) and then the Lagrangian density takes the form
\begin{eqnarray}
S=\int\Phi d^{4}x[-\frac{1}{\kappa}R(\Gamma,g)-m^{4}f(u)+
\frac{1}{2}g^{\mu\nu}\varphi,_{\mu}\varphi,_{\nu}-
V(\varphi)+
\frac{1}{2}\tilde{e}^{2}\varphi^{2}g^{\mu\nu}\tilde{A}_{\mu}\tilde{A}_{\nu}]
 \label{V18}
\end{eqnarray}
where $|\phi|=\frac{1}{2}\varphi$.

The constraint (\ref{II3}) has now the form 
\begin{equation}
-2uf^{\prime} (u)+f(u)+\frac{1}{m^{4}}[V(\varphi)+M]=0.
 \label{V19}
\end{equation}

Varying with respect to $A_{\nu\alpha\beta}$ we get
\begin{equation}
\partial_{\mu}(\chi f^{\prime}\varepsilon^{\mu\nu\alpha\beta})=0
 \label{V20}
\end{equation}
which gives
\begin{equation}
\chi f^{\prime} =\omega
 \label{V21}
\end{equation}
where $\omega$ is a dimensionless integration constant.

Equation for the gauge field $\tilde{A}_{\mu}$ 
\begin{equation}
\frac{1}{\sqrt{-g}}\partial_{\mu}[\chi f^{\prime}\sqrt{-g}
\tilde{F}^{\mu\nu}]
+\frac{\tilde{e}^{2}}{4}\varphi^{2}\chi\tilde{A}^{\nu}
=0
 \label{V22}
\end{equation}
which is apparently nonlinear in $\tilde{F}^{\mu\nu}$ is reduced to the
form 
\begin{equation}
\frac{1}{\sqrt{-g}}\partial_{\mu}[\sqrt{-g}\tilde{F}^{\mu\nu}]
+\frac{\tilde{e}^{2}}{4\omega}\varphi^{2}\chi
g^{\nu\alpha}\tilde{A}_{\alpha}
=0
 \label{Ch1}
\end{equation}
due to Eq. (\ref{V21}).

The gravitational equations originated by the variation of $g^{\mu\nu}$ 
take the form
\begin{eqnarray}
\frac{1}{\kappa}R_{\mu\nu}(\Gamma,g)=
-\frac{1}{2}yf^{\prime}g_{\mu\nu}+
\frac{1}{2}f^{\prime}\left[\tilde{F}^{\alpha\beta}
\tilde{F}_{\alpha\beta}g_{\mu\nu}-
4\tilde{F}_{\mu\alpha}\tilde{F}_{\nu\beta}g^{\alpha\beta}\right]
+\frac{1}{2}\varphi_{,\mu}\varphi_{,\nu}+
\frac{\tilde{e}^{2}}{2}\varphi^{2}\tilde{A}_{\mu}\tilde{A}_{\nu}
 \label{V23}
\end{eqnarray}
after using Eq. (\ref{V1}).

The scalar field equation is
\begin{equation}
(-g)^{-1/2}\partial_{\mu}(\sqrt{-g}g^{\mu\nu}\partial_{\nu}\varphi)
+\sigma,_{\mu}\varphi ^{,\mu}+V^{\prime}(\varphi)+
\tilde{e}^{2}\varphi g^{\alpha\beta}\tilde{A}_{\alpha}\tilde{A}_{\beta}=0,
 \label{Ch2}
\end{equation}

The $\chi$- field enters both in the gravitational Eqs. (\ref{V23})
(through the connection) and in the matter equations  (\ref{Ch1}),
(\ref{Ch2}). In
order to
see easily the physical content of the model, we have to perform a
{\em conformal transformation}
\begin{equation}
\overline{g}_{\mu\nu}(x)=\chi
g_{\mu\nu}(x); \qquad \varphi\rightarrow\varphi
\label{Ch3}
 \end{equation}
 to obtain an "Einstein picture".
Notice that now this transformation is not a symmetry and indeed
changes the form of equations.
After 
 rescaling $\tilde{A}_{\mu}$ and $\tilde{e}^{2}$ to ${A}_{\mu}$ 
and ${e}^{2}$ 
\begin{equation}
A_{\mu}\equiv 2\sqrt{\omega}\tilde{A}_{\mu}; \qquad
e=\frac{\tilde{e}}{2\sqrt{\omega}}
 \label{Ch4}
\end{equation}
 with $\omega >0$, we obtain the resulting equations in the Einstein frame
which are of canonical form for a standard SSB gauge theory in the GR
formalism:
\begin{equation}
\frac{1}{\sqrt{-\overline{g}}}\partial_{\mu}(\sqrt{-\overline{g}}
\enspace\overline{g}^{\mu\nu}\partial_{\nu}\varphi)+
\frac{dV_{eff}}{d\varphi}+
e^{2}\varphi g^{\alpha\beta}A_{\alpha}A_{\beta}
=0
 \label{Ch5}
\end{equation}
\begin{equation}
\frac{1}{\sqrt{-\overline{g}}}\partial_{\mu}(\sqrt{-\overline{g}}
\enspace\overline{g}^{\mu\alpha}\overline{g}^{\nu\beta}
F_{\alpha\beta})+
e^{2}\varphi^{2}\overline{g}^{\nu\alpha}A_{\alpha}
=0
 \label{Ch6}
\end{equation}
\begin{equation}
G_{\mu\nu}(\overline{g}_{\alpha\beta})=\frac{\kappa}{2}T_{\mu\nu}
 \label{Ch7}
\end{equation}
\begin{equation}
T_{\mu\nu}=
\varphi_{,\mu}\varphi_{,\nu}-
\frac{1}{2}g_{\mu\nu}\varphi_{,\alpha}\varphi_{,\beta}
\overline{g}^{\alpha\beta}+V_{eff}(\varphi)\overline{g}_{\mu\nu}+
\frac{1}{4}\overline{g}_{\mu\nu}F_{\alpha\beta}F^{\alpha\beta}
-F_{\mu\alpha}F_{\nu\beta}\overline{g}^{\alpha\beta}+
e^{2}\varphi^{2}(A_{\mu}A_{\nu}-
\frac{1}{2}\overline{g}_{\mu\nu}A_{\alpha}A^{\alpha})
 \label{Ch8}
\end{equation}
Here
\begin{equation}
G_{\mu\nu}(\overline{g}_{\alpha\beta})\equiv
R_{\mu\nu}(\overline{g}_{\alpha\beta})-
\frac{1}{2}\overline{g}_{\mu\nu}R(\overline{g}_{\alpha\beta})
 \label{Ch9}
 \end{equation}
is the Einstein tensor in the Riemannian space-time with metric
$\overline{g}_{\mu\nu}$.

The new effective scalar field potential
appears now in two different forms in Eqs. (\ref{Ch5}) and (\ref{Ch8}):
\begin{equation}
\frac{dV_{eff}}{d\varphi}\equiv
V_{eff}^{\prime}=\frac{1}{\chi}\frac{dV}{d\varphi}=
\frac{1}{\omega}\frac{df}{du}\frac{dV}{d\varphi}
 \label{V24}
\end{equation}
and
\begin{equation}
 V_{eff}(\varphi)=
\frac{y}{\omega}(f^{\prime}(u))^{2}.
 \label{V25}
\end{equation}
Here $y, \ u\equiv y/m^{4}$ and $f(u)$ are functions of $\varphi$ due to 
the constraint (\ref{V19}). It can be shown 
that two different form of appearance of $V_{eff}$ and $V^{\prime}_{eff}$ 
in the gravitational field equations and in the scalar field equation 
correspondingly, are self consistent (the reason for this is the existence 
of Bianchi identities). This consistency also may be shown by taking the 
derivative of Eq. (\ref{V25}) with respect to $\varphi$ and using the 
derivative of the constraint (\ref{V19}) with respect to $\varphi$. As the 
result we obtain Eq. (\ref{V24}). 

Looking at Eq. (\ref{V24}) we see that there are two ways of obtaining an 
extremum of $V_{eff}(\varphi)$:

 (a) The first one is when 
$\frac{dV}{d\varphi}=0$ which corresponds to an extremum of the original 
potential $V(\varphi)$. In this case there is no reason for the vanishing 
$V_{eff}(\varphi)$ in such an extremum if we do not resort to some kind 
unnatural fine tuning.

 (b) The second way is to consider the situation where 
\begin{equation}
\frac{df}{du}|_{u=u_{0}\not=0}=0
 \label{V26}
\end{equation} 
and the appropriate value of $\varphi_{0}$ is related to $u_{0}$ by the 
constraint (\ref{V19}). For this extremum of the effective potential we 
see immediately from Eq. (\ref{V25}) that $V_{eff}(\varphi_{0})=0$ 
without any assumption about $\frac{dV}{d\varphi}$, that is without fine 
tuning. 

If we assume that $y_{0}=\frac{u_{0}}{m^{4}}$ is positive then it is 
clear also from  (\ref{V25}) that $\varphi_{0}$ (where 
both $V_{eff}(\varphi_{0})=0$ and 
$\frac{dV_{eff}}{d\varphi}|_{\varphi=\varphi_{0}}=0$), is a 
minimum since any small fluctuations bring us to a higher positive value of 
$V_{eff}$. In this case the vacuum is defined both by value of the gauge 
complex condensate $u=u_{0}$ and by the scalar condensate $\varphi 
=\varphi_{0}$ satisfying the condition
\begin{equation}
f(u_{0})+\frac{1}{m^{4}}[V(\varphi_{0})+M]=0.
 \label{V27}
\end{equation}
which follows from the constraint (\ref{V19}) and Eq.(\ref{V26}). Notice 
that transition to the Einstein frame does not change the value of the 
gauge complex condensate $u_{0}$ since it is defined by the value of 
$\varphi_{0}$ due to Eq. (\ref{V27}). 
 
It is very important to notice that Eq. (\ref{V27}) represents {\em the 
exact mutual cancellation of the contributions to the vacuum energy of 
the integration constant $M$, the scalar field condensate and 
the gauge complex condensate}.

It can be shown explicitly (by using the constraint (\ref{V19}), its 
derivative and Eq. (\ref{V24}) that the mass square of the scalar 
particle is
\begin{equation}
V^{\prime\prime}_{eff}(\varphi_{0})=
\frac{1}{2\omega y_{0}}[V^{\prime}(\varphi_{0})]^{2}.
 \label{V28}
\end{equation}  
which is positive if both $\omega >0$ and $y_{0}>0$. In this case the 
state $y=y_{0}$, $y=y_{0}$ is classically stable and it has zero 
energy density. In this state, the vector boson $A_{\mu}$ acquires mass
\begin{equation}
m_{A}^{2}=e^{2}\varphi_{0}^{2}
\label{mA}
\end{equation}

As it has been noted in Introduction, equations of motion derived from  
the action (\ref{Action}) are
invariant under the global dilatation of the measure fields $\varphi_{a}$:
\begin{equation}
\varphi_{a}\rightarrow\alpha_{a}\varphi_{a}, \qquad \alpha_{a}=const
\label{D1}
\end{equation}
or in terms of the measure $\Phi$
\begin{equation}
\Phi\rightarrow\alpha\Phi, \qquad \alpha=\prod \alpha_{a}
\label{D2}
\end{equation}

In the Einstein frame (see Eq. (\ref{Ch3})) this symmetry takes the form
of a global space-time dilatation
\begin{equation}
\Phi(x)\rightarrow\alpha\Phi(x) ;\qquad
\overline{g}_{\mu\nu}(x)\rightarrow
\alpha\overline{g}_{\mu\nu}(x) ;\qquad 
(\chi(x)\rightarrow\alpha\chi(x))
\label{D3}
 \end{equation}

The appearance of the gauge complex condensate (\ref{V21}) parameterized by
the integration constant $\omega$, spontaneously breaks this symmetry. As
we can see from Eqs. (\ref{Ch4}), (\ref{mA}) and (\ref{V28}) all coupling
constants and masses are defined by the dilatation breaking integration
constant $\omega$. 

One can see that if $y>0$ in the extremum where $\frac{dV}{d\varphi}=0$, 
then this state can serve as 
a phase with positive effective cosmological constant and therefore 
inflation becomes possible as well. This vacuum is smoothly connected by 
dynamical evolution of the scalar field $\varphi$ with the zero energy 
density vacuum $y=y_{0}$, $y=y_{0}$, thus providing a way to achieve 
inflation and transition (after a standard reheating period) to a $\Lambda 
=0$ phase without fine tuning.

\section{Models satisfying the weak NGVE principle}

\subsection{General features of the weak NGVE theory}

As we have seen in Introduction and in Sec. IIA, an action of the form
(\ref{Action}) is invariant (up to an integral of a total divergence)
under the SGMF transformations (\ref{SG}). However, the existence of such
symmetry is not affected by the existence of an additional contribution to
the action of the form $S_{2}$ in Eq. (\ref{Action12}). In this case the
SGMF symmetry is modified as in Eq. (\ref{SG1}). In this section we want
to study this more general structure. Notice that in the weak NGVE theory,
the invariance of classical equations of motion under the global dilaton
transformations (\ref{D2}) is explicitly violated.

We are starting now from Eq. (\ref{Action12}) in the case when neither
$L_{1}$
nor $L_{2}$
contain the measure fields $\varphi_{a}$ and where for simplicity we
assume that no $\Gamma^{\lambda}_{\mu\nu}$ - dependence enters in
$L_{2}\equiv L_{m2}=L_{m2}(g_{\mu\nu}, matter fields)$ and $L_{1}$ has the
form $L_{1}=-\frac{1}{\kappa}R(g,\Gamma)+L_{m1}(g_{\mu\nu}, matter
fields)$.

Variation with respect to the measure fields $\varphi_{a}$ leads to the
equation 
\begin{equation}
A^{\mu}_{a}\partial_{\mu}\lbrack -\frac{1}{\kappa}R(\Gamma,g)+
L_{m1}\rbrack =0
\label{FEM1}
\end{equation}

Just as in Sec. IIA, if 
 $\Phi\neq 0$, it follows from Eq. (\ref{FEM1}) 
\begin{equation}
 -\frac{1}{\kappa}R(\Gamma,g)+
L_{m1}=M=constant
\label{w1}
\end{equation}

Varying with respect to $g^{\mu\nu}$ we get
\begin{equation}
\Phi\left(-\frac{1}{\kappa}R_{\mu\nu}(\Gamma)+
\frac{\partial L_{m1}}{\partial g^{\mu\nu}}\right)-
\frac{1}{2}\sqrt{-g}L_{m2}g_{\mu\nu}+
\sqrt{-g}\frac{\partial L_{m2}}{\partial g^{\mu\nu}}=0
\label{Gw}
\end{equation}

The consistency condition of Eqs. (\ref{w1}) and (\ref{Gw})  is the
constraint
\begin{equation}
g^{\mu\nu}\left(\frac{\partial L_{m1}}{\partial g^{\mu\nu}}+
\frac{1}{\chi}\frac{\partial L_{m2}}{\partial g^{\mu\nu}}\right)-
\left(L_{m1}+\frac{2}{\chi}L_{m2}-M\right)=0
\label{wc}
\end{equation}
which takes place of the constraint (\ref{II3}) of Sec. IIA. Notice that
constraint (\ref{wc}) is satisfied automatically (that is without using
equations of motions) in the case when $L_{m1}$ and
$L_{m2}$ are homogeneous functions of $g^{\mu\nu}$ of degree 1 and 2
correspondingly. In such a case the theory possesses the LES
(\ref{ES1}), (\ref{ES2}) and then, similarly to the strong NGVE theory
(see the end of Sec.IIB), the $\chi$ contributions into equations of
motion can be
eliminated by the transformation (\ref{ES1}), (\ref{ES2}) with $J=\chi$.

\subsection{The simplest model}

To demonstrate how the theory works when it is based on the weak NGVE
principle, we start here from the
simplest 
model \cite{PASCOS}, \cite{Fried98}  including
scalar field
$\varphi$ and gravity
according to the prescription of the NGVE principle and in addition to 
this we include the standard cosmological constant term. 
So, we consider an action 
\begin{equation}
S=\int L_{1}\Phi d^{4}x + \int\Lambda\sqrt{-g}d^{4}x
\label{AW}
\end{equation}
where 
$L_{1}=-\frac{1}{\kappa}R(\Gamma,g)+
\frac{1}{2}g^{\mu\nu}\varphi_{,\mu}\varphi_{,\nu}
-V(\varphi)$ and $\Lambda =const.$. Notice that now the global dilatation
symmetry (\ref{D2}) is
explicitly broken.

Performing the variation with respect to the measure fields $\varphi_{a}$ 
 we obtain equations 
$A^{\mu}_{a}\partial_{\mu}L_{1}=0$
where $A^{\mu}_{b}$ is given by Eq. (\ref{DefA}). 
If $\Phi\neq 0$, 
then it follows from the last equations that $L_{1}=M=const$.

Varying the action (\ref{AW}) with respect to $g^{\mu\nu}$ we get
\begin{equation}
\Phi(-\frac{1}{\kappa}R_{\mu\nu}(\Gamma)+
\frac{1}{2}\varphi_{,\mu}\varphi_{,\nu})
-\frac{1}{2}\sqrt{-g}\Lambda g_{\mu\nu}=0
\label{GW}
\end{equation}

Contracting Eq.(\ref{GW}) with $g^{\mu\nu}$ and using equation $L_{1}=M$
we 
obtain the constraint
\begin{equation}
M+V(\varphi)-\frac{2\Lambda}{\chi}=0
\label{CW}
\end{equation}
where again $\chi\equiv\Phi/\sqrt{-g}$.

The scalar field $\varphi$ equation has the same form as Eq. (\ref{SE}).

The derivatives of the field $\sigma$ enter both the gravitational 
equation (\ref{GW}) (through the connection) and in the scalar field
equation.
 By a conformal transformation (\ref{Ch3})
to an "Einstein picture" we get, after using the constraint (\ref{CW}),
the 
canonical form of equation for the scalar field  
and of the gravitational equations in the Riemannian
space-time with metric 
$\overline{g}_{\alpha\beta}$
\begin{equation}
\frac{1}{\sqrt{-\overline{g}}}\partial_{\mu}(\sqrt{-\overline{g}}
\enspace\overline{g}^{\mu\nu}\partial_{\nu}\varphi)+
\frac{dV_{eff}}{d\varphi}=0
 \label{sw1}
\end{equation}
\begin{equation}
R_{\mu\nu}(\overline{g}_{\alpha\beta})-
\frac{1}{2}\overline{g}_{\mu\nu}R(\overline{g}_{\alpha\beta})
=\frac{\kappa}{2}T^{eff}_{\mu\nu}(\varphi)
\label{gw}
\end{equation}
where 
\begin{equation}
T^{eff}_{\mu\nu}(\varphi)=\varphi_{,\mu}\varphi_{,\nu}-
\frac{1}{2}\overline{g}_{\mu\nu}
\varphi_{,\alpha}\varphi_{,\beta}\overline{g}^{\alpha\beta}+
\overline{g}_{\mu\nu}V_{eff}(\varphi)
\label{TW}
\end{equation}
and
\begin{eqnarray}
\frac{dV_{eff}}{d\varphi} & = &\frac{1}{\chi}\frac{dV}{d\varphi} \\
V_{eff}(\varphi) & = & 
\frac{1}{4\Lambda}[M+V(\varphi)]^{2}
\label{VW}
\end{eqnarray}

We see that for any analytic function $V(\varphi)$, the effective
potential 
in the Einstein picture has an extremum, i.e. $V_{eff}^{\prime}=0$, 
either when $V^{\prime}=0$ or $V+M=0$. The extremum $\varphi=\varphi_{1}$
where 
$V^{\prime}(\varphi_{1})=0$ has nonzero energy density 
$[M+V(\varphi_{1})]^{2}/4\Lambda$ if a fine tuning is not assumed. In 
contrast to this, if $\Lambda >0$, the state $\varphi =\varphi_{0}$ where 
$V(\varphi_{0})+M=0$ is the absolute minimum and therefore $\varphi_{0}$
is a 
true vacuum with zero cosmological constant without any fine tuning. A mass 
square of the scalar field describing small fluctuations around 
$\varphi_{0}$ is
$m^{2}=\frac{1}{2\Lambda}[V^{\prime}(\varphi_{0})]^{2}$.
Exploiting the possibility to choose any analytic $V(\varphi)$, we can
pick 
the structure of $V_{eff}(\varphi)$ so that it
allows for an
inflationary era, the possibility of reheating after scalar field
oscillations and the setting down to a zero cosmological constant phase
at the later stages of cosmological evolution, without fine tuning.

Notice that if $V+M$ reaches the value zero at some value of $\varphi
=\varphi_{0}$, this point represents the absolute minimum of the effective
potential. If this is not the case for a particular choice of potential
$V$
and of integration constant $M$, it is
always possible to choose an infinite range of values of $M$ where this
will happen. Therefore no fine tuning of parameters has to be invoked,
the zero value of the true vacuum energy density appears naturally in this
theory. 

It is interesting that in the TVS $\chi\rightarrow\infty$, that is the
dynamical measure $\Phi$ dominates the measure $\sqrt{-g}$. Therefore, in
the TVS we
observe a dominance of the first term (even if $\Lambda$ is very big) in
Eq. (\ref{AW}) which implies the restoration of the global dilatation
symmetry (\ref{D2}) in the TVS.

As an example let us consider the model with the quadratic potential
$V(\varphi)=\frac{\mu^{2}}{2}\varphi^{2}$, which in the standard GR is
associated with massive non selfinteracting real scalar field theory. In
the
theory under consideration, however, we obtain a very different result:
the effective potential in the Einstein picture appears
\begin{equation}
V_{eff}=\frac{\mu^{4}}{16\Lambda}(\varphi^{2}-\varphi_{0}^{2})^{2}, \qquad
\varphi_{0}^{2}=-\frac{2M}{\mu^{2}}
\label{q}
\end{equation}
Notice that it follows from the constraint (\ref{CW}) that
$M=\frac{2\Lambda}{\chi} -
\frac{1}{2}\mu^{2}\varphi^{2}\rightarrow -\frac{1}{2}\mu^{2}\varphi^{2}$
as
$\varphi$ approaches the TVS (that is $\chi^{-1}\rightarrow 0$).
Therefore,
if
$\mu^{2}>0$, then for the TVS to exist, the integration constant $M$ must
be $M\leq 0$.

We see that for $\Lambda >0$, spontaneous breaking of the discrete
symmetry
$\varphi\rightarrow -\varphi$ takes place if $M\mu^{2}<0$. Yet, the vacuum
energy at the absolute minimum $\varphi=\pm |\varphi_{0}|$ is zero without
fine
tuning. Furthermore, the mass of the scalar field is
$m^{2}=\frac{\mu^{4}}{2\Lambda}\varphi_{0}^{2}=-\frac{M\mu^{2}}{\Lambda}$
which as we see depends on the integration constant $M$. The mass $m$ is
therefore a "floating" physical parameter, since $M$ does not appear in
the original Lagrangian but it is determined by initial conditions of the
Universe.

If $\varphi$ is replaced by a complex field $\phi$ and $\varphi^{2}$ by
$\phi^{\ast}\phi$, we obtain the SSB of a continuous symmetry with
standard consequences. In addition, a model of cosmology that can include
an inflationary phase taking place in a false vacuum and transition to a
zero cosmological constant phase is obtained without fine tuning.

Further generalizations do not modify the qualitative nature of the
effects described here and they will be studied elsewhere.
For example, considering
a term of the form $\int U(\varphi)\sqrt{-g}d^{4}x$ instead of 
$\Lambda\int \sqrt{-g}d^{4}x$ in Eq.(\ref{AW}),
one can see that the resulting effective potential vanishes when 
$V(\varphi_{0})+M=0$ if $U(\varphi_{0})\neq 0$ since it equals 
$V_{eff}=\frac{1}{4U(\varphi)}(V+M)^{2}$. 

A model based on the weak NGVE principle which incorporates global scale
invariance is considered in Refs. \cite{G1} and \cite{G11}.

\subsection{Including the gauge fields}

The weak NGVE principle allows to incorporate  gauge fields in a way
which is  simpler  than in the context of the strong NGVE
principle. Taking in Eqs. (\ref{Action12})
\begin{eqnarray}
L_{1} & = & -\frac{1}{\kappa}R(\Gamma, g)+g^{\mu\nu}(\partial_{\mu} 
-ieA_{\mu})\phi (\partial_{\nu}+ieA_{\nu})\phi^{*}-V(\phi)
\nonumber\\
L_{2} & = &
-\frac{1}{4}g^{\alpha\beta}g^{\mu\nu}F_{\alpha\mu}F_{\beta\nu}+\Lambda
\label{LW}
\end{eqnarray}
we see that except of the $V$-term in $L_{1}$ and $\Lambda$-term in 
$L_{2}$, the action (\ref{Action12}) is invariant under the LES (notice
that in $S_{2}$ the LES 
takes the form of the conformal transformation  of the metric
$g_{\mu\nu}(x)=J^{-1}(x)g^{\prime}_{\mu\nu}(x)$). Therefore, as one can
check explicitly, the gauge
field does not enter in the constraint which turns out to be identical to  
 Eq. (\ref{CW}). 

For the gauge field in the unitary gauge we get 
\begin{equation}
(-g)^{-1/2}\partial_{\mu}(\sqrt{-g}g^{\mu\alpha}g^{\nu\beta}F_{\alpha\beta})
+ e^{2}\chi\varphi^{2}g^{\mu\nu}A_{\mu}=0
\label{WGE}
\end{equation}
 In the Einstein
conformal frame where the metric is $\overline{g}_{\mu\nu}=\chi
g_{\mu\nu}$, the gauge field equation takes the canonical form
\begin{equation}
\frac{1}{\sqrt{-\overline{g}}}
\partial_{\mu}(\sqrt{-\overline{g}}\enspace\overline{g}^{\mu\alpha}
\overline{g}^{\nu\beta}F_{\alpha\beta})+e^{2}\varphi^{2}
\overline{g}^{\mu\nu}A_{\mu}=0
\label{GE}
\end{equation}

The gravitational and scalar field equations including the effective
scalar field potential in the Einstein picture
coincide with  the appropriate equations of the model in the previous
Subsection IIIB.  Keeping the assumption that 
$\Lambda >0$ we can see that all conclusions
concerning the vanishing of the
vacuum energy in the TVS $\varphi =\varphi_{0}$ where
$V(\varphi_{0})+M=0$ remain unchanged. In addition to this, the Higgs
mechanism for the mass generation in the true vacuum state $\varphi_{0}$
works now in a regular way in contrast to the model based on the strong
NGVE principle where the effective coupling constant (and as a
consequence, the mass of the gauge boson) depends on the integration
constant.  

 \bigskip
\section{The inclusion of fermions}

\bigskip
\subsection{Fermions in the NGVE theory}

\bigskip

To present a complete enough picture let us consider a strong NGVE model
including 
gravity, gauge and scalar field sectors (as in the action (\ref{V18}), 
once again in the unitary gauge) and, in addition, the fermionic sector 
(for notations see Appendixes E and F):
\begin{eqnarray}
S=\int\Phi d^{4}x[-\frac{1}{\kappa}V^{a\mu}V^{b\nu}R_{\mu\nu 
ab}(\omega)-m^{4}f(u) +\frac{1}{2}g^{\mu\nu}\varphi,_{\mu}\varphi,_{\nu}-
V(\varphi)+
\frac{1}{2}\tilde{e}^{2}\varphi^{2}g^{\mu\nu}\tilde{A}_{\mu}\tilde{A}_{\nu}
\nonumber\\
+\frac{i}{2}\overline{\Psi}\left\{\gamma^{a}V_{a}^{\mu}
(\overrightarrow{\partial}_{\mu}+\frac{1}{2}\omega_{\mu}^{cd}\sigma_{cd}-
i\tilde{e}\tilde{A}_{\mu})
-(\overleftarrow{\partial}_{\mu}-\frac{1}{2}\omega_{\mu}^{cd}\sigma_{cd}+
i\tilde{e}\tilde{A}_{\mu})
\gamma^{a}V_{a}^{\mu}\right\}\Psi+U(\overline{\Psi}\Psi)]
 \label{VI1}
 \end{eqnarray}
where the selfinteraction term $U(\overline{\Psi}\Psi)$ depending on the 
argument $\overline{\Psi}\Psi$ remains unspecified in this subsection.

Equation for $\Psi$ which follows from the action (\ref{VI1}) is
\begin{equation}
\left\{i\left [V_{a}^{\mu}\gamma^{a}\left 
(\partial_{\mu}-i\tilde{e}\tilde{A}_{\mu}\right )+
\gamma^{a}C^{b}_{ab}
+\frac{1}{4}\omega_{\mu}^{cd}(\gamma^{a}\sigma_{cd}+
                              \sigma_{cd}\gamma^{a})V^{\mu}_{a}\right ]+
\frac{1}{2}\gamma^{a}V^{\mu}_{a}\sigma_{,\mu}
+U^{\prime}(\overline{\Psi}\Psi)\right\}\Psi=0
 \label{VI2}
 \end{equation}
where 
\begin{equation}
C^{b}_{ab}=\frac{1}{2\sqrt{-g}}
\partial_{\mu}\left (\sqrt{-g}V^{\mu}_{a}\right )
 \label{VI3}
\end{equation}
is the trace of the so-called Ricci rotation coefficients \cite{Gasp} and 
$U^{\prime}$ is derivative of $U$ with respect to its argument 
$\overline{\Psi}\Psi$. Spin-connection $\omega_{\mu}^{cd}$ is defined by 
Eqs. (\ref{C4})- (\ref{C7}). Remind that $\sigma$-field 
is defined by Eq. (\ref{ski}).

After the transition to the Einstein frame by means of the conformal 
transformations
\begin{equation}
V^{\prime}_{a\mu}(x)=\chi^{1/2}(x)V_{a\mu}(x)
 \label{VI4}
\end{equation}
\begin{equation}
\Psi^{\prime} (x)=\chi^{-1/4}(x)\Psi (x); \ 
\overline{\Psi}^{\prime} (x)=\chi^{-1/4}(x)\overline{\Psi} (x)
 \label{VI5}
\end{equation}
\begin{equation}
\varphi\rightarrow\varphi ; \ \tilde{A}_{\mu}\rightarrow\tilde{A}_{\mu}
 \label{VI6}
\end{equation}
and performing the rescalings (\ref{Ch4}), Eq. (\ref{VI2}) is reduced to
the 
form
\begin{equation}
\left\{i\left [V_{a}^{\prime\mu}\gamma^{a}\left
(\partial_{\mu}-ieA_{\mu}\right )+
\gamma^{a}C^{\prime b}_{ab}
+\frac{i}{4}\omega_{\mu}^{\prime cd}\varepsilon_{abcd}\gamma^{5}\gamma^{b}
V^{a\mu}\right ]
+\chi^{-1/2}U^{\prime}(\overline{\Psi}^{\prime}\Psi^{\prime})\right\}
\Psi^{\prime}=0
 \label{VI7}
 \end{equation}
where $U^{\prime}(\overline{\Psi}^{\prime}\Psi^{\prime})=
U^{\prime}(\overline{\Psi}\Psi)|_{\Psi=\chi^{1/4}\Psi^{\prime};
\overline{\Psi}=\chi^{1/4}\overline{\Psi}^{\prime}}$ and $\chi$ enters 
only as a factor in front of 
$U^{\prime}(\overline{\Psi}^{\prime}\Psi^{\prime})$. In Eq. (\ref{VI7}), 
\begin{equation}
C^{\prime b}_{ab}=\frac{1}{2\sqrt{-g^{\prime}}}
\partial_{\mu}\left (\sqrt{-g^{\prime}}V^{\prime\mu}_{a}\right )
 \label{VI8}
 \end{equation}
and the spin-connection $\omega_{\mu}^{\prime cd}$ is
\begin{equation}
\omega_{\mu}^{\prime cd}=\omega_{\mu}^{cd}(V^{\prime\alpha}_{a})+
K_{\mu}^{cd}(V^{\prime\alpha}_{a},\Psi^{\prime},\overline{\Psi}^{\prime})
 \label{VI9}
 \end{equation}
Notice that after the conformal 
transformation (\ref{VI4})-(\ref{VI6}), the $\sigma$-contribution to the 
spin-connection (the second term in the r.h.s. of Eq. (\ref{C4})) is 
canceled. 

Notice also that once again we performed the rescaling (\ref{Ch4}) in 
order to provide the standard form of the appropriate equations for the 
gauge and gravitational fields.

\bigskip
\subsection{Nambu - Jona-Lasinio model in the NGVE theory}

\bigskip

As it is shown in Appendix E, the famous Nambu - Jona-Lasinio (NJL) model 
\cite{NJL} works in a special way in the context of the NGVE theory: in 
the theory with only fermionic matter, the constraint does not impose 
restrictions on dynamics (see also \cite{GK2}).

If we generalize the model towards a realistic theory by the 
consideration both bosonic and fermionic sectors as in the previous 
subsection, then still the fermionic part does not contribute to the 
constraint if we restrict ourselves to the NJL - type model.
In this case the constraint works exactly as in Sec. III where bosonic 
sector breaks the LES explicitly and the constraint becomes an equation 
for $\chi$. The modification of the NJL type model for the case of the
weak NGVE theory (see Eq. (\ref{Action12})) is evident if the fermion
sector enter only in $S_{1}$.

In the Einstein frame, the NJL term (see the last term 
in Eq. (\ref{VI7})) remains its original form in terms of the transformed 
fields ${\Psi}^{\prime}$ and $\overline{\Psi}^{\prime}$. 

As it is well known \cite{Nambu-Bardeen}, the the NJL mechanism 
allows for a dynamical mass generation in realistic models. Since in the 
Einstein frame the 
theory takes the canonical form, therefore the same mechanisms of the 
dynamical mass generation can be applied also here.

\bigskip
\subsection{Classical model for generating mass of fermions}

\bigskip 

As we have discussed in the previous subsection, there is a possibility 
to generate masses of fermions in the NGVE theory due to the quantum 
effect in the NJL model. Now we are going to show that the NGVE theory in 
the context of the model with the action (\ref{VI1}) allows for a purely 
classical mechanism of obtaining fermion masses.

The constraint (\ref{B7}) corresponding to the model (\ref{VI1}) is
\begin{equation}
\overline{\Psi}\Psi U^{\prime}(\overline{\Psi}\Psi)-
2U(\overline{\Psi}\Psi)-2\left [m^{4}(2uf^{\prime}(u)-f(u))-
(V(\varphi)+M)\right ]=0
 \label{VI10}
 \end{equation}   
where Eq. (\ref{B10}) has been used.

Directing our attention to the situation where the scalar field 
excitations around the vacuum \{$u=u_{0}$, $\varphi=\varphi_{0}$ with 
condition (\ref{V19})\}, (see Sec.IID) are ignored, we get the 
constraint for the first order fluctuations of $\Psi$ and $\overline{u}
\equiv u-u_{0}$
\begin{equation}
\overline{\Psi}\Psi U^{\prime}(\overline{\Psi}\Psi)-
2U(\overline{\Psi}\Psi)-
4m^{4}u_{0}f^{\prime\prime}(u_{0})\,\overline{u}=0
 \label{VI11}
 \end{equation}
where Eq. (\ref{V26}) has been used.

We consider now the  model with  fermionic selfinteraction of the form
\begin{equation}
U(\overline{\Psi}\Psi)=-C(\overline{\Psi}\Psi)^{q}, \qquad C=km^{4-3q}
 \label{VI12}
 \end{equation}
where $k>0$ is a dimensionless parameter. 

Remembering that our aim is to study the theory in the Einstein frame, we 
first calculate $\chi$. From Eq. (\ref{V21}) we have $\chi=
\frac{\omega}{f^{\prime}}$ and expanding $f^{\prime}$ around $u=u_{0}$ we get
$f^{\prime}=f^{\prime}(u_{0})+f^{\prime\prime}(u_{0})\overline{u}+\ldots =
f^{\prime\prime}(u_{0})\overline{u}$. $\overline{u}$ can be obtained from 
Eq. (\ref{VI11}) and for special form (\ref{VI12}) we obtain, after a 
simple computation, the following equation for $\chi$
\begin{equation}
\frac{\omega}{\chi}=\frac{C}{4u_{0}m^{4}}\left 
[(2-q)\chi^{q/2}(\overline{\Psi}^{\prime}\Psi^{\prime})^{q}\right ]
 \label{VI13}
 \end{equation}
in terms of the fermion field $\Psi^{\prime}$ in the Einstein frame (see 
Eq. (\ref{VI5})), that is
\begin{equation}
\chi=\left (\frac{C(2-q)}{4\omega u_{0}m^{4}}\right )^{-\frac{2}{q+2}}
\left (\overline{\Psi}^{\prime}\Psi^{\prime}\right )^{-\frac{2q}{q+2}}
 \label{VI14}
 \end{equation}
 
Finally, we observe that in the Einstein frame, the last term in Eq. 
(\ref{VI7}) has the form of a mass term, with a mass $m_{f}$ given by
\begin{equation}
m_{f}=Cq\chi^{-1/2}
 (\overline{\Psi}\Psi)^{q-1}=Cq\chi^{\frac{q-2}{2}}
(\overline{\Psi}^{\prime}\Psi^{\prime})^{q-1}
 \label{VI15}
 \end{equation}

Such a term will be a legitimate mass term only if $m_{f}$ as given by 
Eq. (\ref{VI15}) is a genuine constant. From Eqs. (\ref{VI14}) and 
(\ref{VI15}) we obtain that $m_{f}$ is given by 
\begin{equation}
m_{f}=Cq\left (\frac{C(2-q)}{4\omega u_{0}m^{4}}\right )^{\frac{2-q}{q+2}}
(\overline{\Psi}^{\prime}\Psi^{\prime})^{\frac{3q-2}{q+2}}
 \label{VI16}
 \end{equation}

We observe therefore that $m_{f}$ is indeed a genuine constant if 
$q=2/3$:
\begin{equation}
m_{f}=\frac{mk^{3/2}}{\sqrt{3\omega u_{0}}}
 \label{VI17}
 \end{equation}

We conclude therefore that if we start with a  fermion selfinteraction 
term in the original Lagrangian density (see Eq. (\ref{VI1})) of the form
\begin{equation}
U(\overline{\Psi}\Psi)=-C(\overline{\Psi}\Psi)^{2/3},
\label{VI18}
 \end{equation}
 we obtain normal 
propagation of a massive fermion in the (physical) Einstein frame.

If we had taken the normal fermion mass term in the original frame, that
is $q=1$, then in the absence of fermionic condensate we would get zero
fermion mass in the Einstein frame.

One should notice at this point that a selfinteraction of the form 
(\ref{VI18})
 has a remarkable feature: such a term 
comes in the action with a coupling constant $C=km^{2}$ with 
dimensionality $mass^{2}$, just as we are used in the case of bosonic 
masses, like in the case of a vector meson mass term for example.

\bigskip
\section{Unified gauge theories in the context of NGVE theory}

\bigskip 

Now we able to formulate a realistic gauge theory. For the illustration 
of this we take $SU(2)\times U(1)$ model of electroweak interaction. 
However, there are no obstacles to formulate QCD, GUT or any other 
spontaneously broken gauge model in the context of the 
NGVE theory. The common feature of such models in the NGVE theory is that 
{\em the spontaneous symmetry breaking (SSB) in the vacuum 
$\{\varphi_{0},y_{0}\}$  does not 
generate the vacuum energy and therefore the cosmological constant is 
equal to zero in this vacuum}. We present here the model realized in the
context of the strong NGVE theory with the use of the 4-index field
strength. However, unified gauge theories can be formulated without the
4-index field strength following the guide-line of the Sec. III, that is in
the context of the weak NGVE theory.

The $SU(2)\times U(1)$ model of electroweak interaction in the NGVE 
theory has to be considered together with  gravitational interaction and 
the Lagrangian density is the following
\begin{eqnarray}
L=-\frac{1}{\kappa}V^{a\mu}V^{b\nu}R_{\mu\nu 
ab}(\omega)-m^{4}f(u)+
i\,\overline{L}\not\!\!DL+i\,\overline{e}_{R}\not\!\!De_{R}+
i\,\overline{\nu}_{R}\not\!\!D\nu_{R}
\nonumber\\
+|D_{\mu}\varphi|^{2}-V(|\varphi|)
-\lambda_{e}m^{4/3}(\overline{L}\,e_{R}\,\varphi+h.c.)^{2/3}
\label{VII18}
 \end{eqnarray} 
where we choosed the classical model for generating mass of fermions (see
Sec. IVC).

In Eq. (\ref{VII18}) we used notations of Ref. \cite{Oku}:
$SU(2)$ vector gauge field 
$\tilde{A}_{i\mu}$, ($i=1,2,3$.);
\quad $U(1)$ abelian gauge field $\tilde{B}_{\mu}$;\quad Left lepton 
dublet \begin{displaymath}
L=\left( \begin{array}{c}
\nu_{L}\\
e_{L}
\end{array} \right);
\end{displaymath}
Right $SU(2)$ singlets $\nu_{R}$ and $e_{R}$;
$SU(2)$ scalar fields dublet
\begin{displaymath}
\varphi=\left( \begin{array}{c}
\varphi^{+}\\
\varphi^{0}
\end{array} \right)
\end{displaymath}
and corresponding antiparticles $\varphi^{\dag}=(\varphi^{-}, 
\tilde{\varphi^{0}})$. The 
left and right components of 
fermions are defined by $\Psi_{L}\equiv\frac{1}{2}(1+\gamma_{5})\Psi$ and
$\Psi_{L}\equiv\frac{1}{2}(1-\gamma_{5})\Psi$ correspondingly;
$D_{\mu}=\partial_{\mu}-igT_{i} \tilde{A}_{i\mu}-
ig^{\prime}\frac{Y}{2}\tilde{B}_{\mu}$ where the hypercharge $Y$ is:
$Y=-1$ for $\nu_{L}$ and $e_{L}$; \quad  $Y=-2$ for $e_{R}$.\quad  
$T_{i}=\frac{1}{2}\tau_{i}$\quad  for \quad $\varphi$ and $L$. For 
isoscalar fields \, $e_{R}$ \quad and\quad $\nu_{R}$,\quad $T=0$. 
The last term in (\ref{VII18}) is written in the form which provides us 
the mechanism for the fermion mass generation described in Sec.IVC and 
accompanied by SSB. Parameter $\lambda_{e}$ in the last term of Eq. 
(\ref{VII18}) is
dimensionless coupling constant.

Operator $\not\!\!D$ in the third term $\overline{L}\not\!\!DL$ of Eq. 
(\ref{VII18}) is defined as follows:
\begin{equation}
\not\!\!D\equiv\overrightarrow{\not\!\!D}_{L}-\overleftarrow{\not\!\!D}_{L}
\label{VII19}
 \end{equation}
where
\begin{equation}
\overrightarrow{\not\!\!D}_{L}\equiv
\frac{1}{2}V^{\mu}_{a}\gamma^{a}\left(\vec{\partial}_{\mu}
+\frac{1}{2}\omega_{\mu}^{cd}I\sigma_{cd}-
\frac{i}{2}\tilde{g}\tau_{i}\tilde{A}_{i\mu}
+\frac{i}{2}\tilde{g}^{\prime}\tilde{B}_{\mu}\right)
\label{VII20}
 \end{equation}
\begin{equation}
\overleftarrow{\not\!\!D}_{L}\equiv\frac{1}{2}
\left(\overleftarrow{\partial}_{\mu}
-\frac{1}{2}\omega_{\mu}^{cd}I\sigma_{cd}+
\frac{i}{2}\,\tilde{g}\tau_{i}\tilde{A}_{i\mu}
-\frac{i}{2}\tilde{g}^{\prime}\tilde{B}_{\mu}\right)\gamma^{a}V_{a}^{\mu}
\label{VII21}
 \end{equation}
where $I$ is $2\times 2$ unit matrix in the isospin space.
The forth and the fifth terms in Eq. (\ref{VII18}) are defined by equations:
\begin{equation}
\overline{e}_{R}\not\!\!De_{R}\equiv\overline{e}_{R}
(\overrightarrow{\not\!\!D}_{eR}-
\overleftarrow{\not\!\!D}_{eR})e_{R}
\label{VII22}
 \end{equation}
\begin{equation}
\overrightarrow{\not\!\!D}_{eR}\equiv\frac{1}{2}V^{\mu}_{a}\gamma^{a}
\left(\vec{\partial}_{\mu}
+\frac{1}{2}\omega_{\mu}^{cd}\sigma_{cd}
+i\tilde{g}^{\prime}\tilde{B}_{\mu}\right)
\label{VII23}
 \end{equation}
\begin{equation}
\overleftarrow{\not\!\!D}_{eR}\equiv\frac{1}{2}
\left(\overleftarrow{\partial}_{\mu}
-\frac{1}{2}\omega_{\mu}^{cd}\sigma_{cd}+
-i\tilde{g}^{\prime}\tilde{B}_{\mu}\right)\gamma^{a}V_{a}^{\mu}
\label{VII24}
 \end{equation}
\begin{equation}
\overline{\nu}_{R}\not\!\!D\nu_{R}\equiv
\overline{\nu}_{R}(\overrightarrow{\not\!\!D}_{\nu R}-
\overleftarrow{\not\!\!D}_{\nu R})\nu_{R}
\label{VII25}
 \end{equation}
\begin{equation}
\overrightarrow{\not\!\!D}_{\nu R}\equiv\frac{1}{2}V^{\mu}_{a}\gamma^{a}
\left(\vec{\partial}_{\mu}
+\frac{1}{2}\omega_{\mu}^{cd}\sigma_{cd}\right)
\label{VII26}
 \end{equation}
\begin{equation}
\overleftarrow{\not\!\!D}_{\nu R}\equiv\frac{1}{2}
\left(\overleftarrow{\partial}_{\mu}
-\frac{1}{2}\omega_{\mu}^{cd}\sigma_{cd}\right)\gamma^{a}V_{a}^{\mu}
\label{VII27}
 \end{equation}

In the kinetic term of the scalar fields in Eq.(\ref{VII18}) the 
notations are the following:
\begin{equation}
|D_{\mu}\varphi|^{2}\equiv 
(D_{\mu}\varphi)^{\dag}_{i}(D_{\nu}\varphi)^{i}g^{\mu\nu}
 \label{VII28}
 \end{equation}
\begin{equation}
D_{\mu}\varphi\equiv
\left(\partial_{\mu}
-\frac{i}{2}\tilde{g}\tau_{i}\tilde{A}_{i\mu}
-\frac{i}{2}\tilde{g}^{\prime}\tilde{B}_{\mu}\right)\varphi
\label{VII29}
 \end{equation}

The gauge complex $u$, which enters into Eq. (\ref{VII18}) as an argument
of the function $f$ in the second term, is defined now as follows:
\begin{equation}
u\equiv\frac{1}{4}\tilde{F}_{\mu\nu}\tilde{F}^{\mu\nu}+
\frac{1}{2}Tr\tilde{G}_{\mu\nu}\tilde{G}^{\mu\nu}+
\frac{m^{2}}{\sqrt{-g}}\varepsilon^{\mu\nu\alpha\beta}\partial_{\mu}
A_{\nu\alpha\beta}
\label{VII30}
 \end{equation}
where $\tilde{F}_{\mu\nu}\equiv\partial_{\mu}\tilde{B}_{\nu}-
                               \partial_{\nu}\tilde{B}_{\mu}$, \ 
$\tilde{G}_{\mu\nu}\equiv\partial_{\mu}\tilde{A}_{\nu}-
                               \partial_{\nu}\tilde{A}_{\mu}-
i\tilde{g}[\tilde{A}_{\mu}\tilde{A}_{\nu}-\tilde{A}_{\nu}\tilde{A}_{\mu}]$
where $\tilde{A}_{\mu}=\tilde{A}_{i\mu}T_{i}$.

>From the action (\ref{VII18}) we will come to equations of motion and the 
constraint similar to what we got before but with the appropriate 
modifications related to  the  $SU(2)\times U(1)$ theory. Nonabelian 
structure of the tensor $\tilde{G}_{\mu\nu}$ does not change the 
constraint since nonlinear term of $\tilde{G}_{\mu\nu}$ has the same 
degree of homogeneity in $g^{\mu\nu}$ as the linear terms. After we will 
get equations similar to Eqs.(\ref{V21}), (\ref{VI10}), (\ref{V24}) and
(\ref{V25}), we see that 
$\frac{dV_{eff}}{d\varphi}=0$ (without a fine tuning of the original 
potential $V(\varphi)$), may be realized when 
$|\varphi|^{2}=\varphi^{2}_{0}=\frac{1}{2}\eta^{2}$ for some vacuum 
expectation value 
\begin{displaymath}
\left<\varphi\right>=\frac{1}{\sqrt{2}}
\left( \begin{array}{c}
0\\
\eta
\end{array} \right)
\end{displaymath}
\label{VII31}
Then the appropriate gauge complex condensate $u_{0}$ is determined by 
Eq. (\ref{V27}) in such a way that the effective cosmological constant 
defined by Eq. (\ref{V25}) is equal to zero.

{\em Around this vacuum} the constraint and equations of motion take the 
standard form. It is important to note a few remarkable features of the 
theory. First, in spite of the fact that more than one gauge vector field 
enter in the gauge complex $u$ (Eq. (\ref{VII30})), the equation similar to 
Eq. (\ref{V21}) again is enough to provide in the Einstein frame, the form 
similar to Eq. (\ref{Ch6}) for all of the gauge field equations. Second,   
once again, in order to provide the canonical form (\ref{Ch6}) of the
gauge field 
equations, we have to perform the rescaling
\begin{eqnarray}
\vec A_{\mu}\equiv 2\sqrt{\omega}\tilde{\vec A}_{\mu}; \qquad
B_{\mu}\equiv 2\sqrt{\omega}\tilde{B}_{\mu};\nonumber\\
g=\frac{\tilde{g}}{2\sqrt{\omega}}; \qquad 
g^{\prime}=\frac{\tilde{g}^{\prime}}{2\sqrt{\omega}}
 \label{VII32}
\end{eqnarray}
where the integration constant $\omega$ appears as the universal parameter.
It is interesting that with this rescaling, where 
$\tilde{g}\tilde{A}_{i\mu}=gA_{i\mu}$, 
$\tilde{g}^{\prime}\tilde{B}_{\mu}=g^{\prime}B_{\mu}$, we obtain 
the canonical form of all of the equations similar to Eqs. (\ref{Ch5}) - 
(\ref{Ch9}) with the appropriate modification to the $SU(2)\times U(1)$ 
theory. From equations similar to Eq. (\ref{Ch6}) we obtain masses of 
vector bosons. After the standard field redefinition from 
$A_{i\mu}$, $B_{\mu}$ to intermediate vector bosons 
$W^{+}_{\mu}$, $W^{-}_{\mu}$, $Z_{\mu}$ and electromagnetic field 
$A_{\mu}$, we obtain the following expressions for their masses:
\begin{equation}
m_{W}=\frac{1}{2}g\eta =\frac{\tilde{g}}{4\sqrt{\omega}}\eta; \qquad
m_{Z}=\frac{1}{2}\overline{g}\eta; \qquad  
m_{A}=0,
\label{VII33}
 \end{equation}
where
\begin{equation}
\overline{g}=\sqrt{g^{2}+g^{\prime 2}} = 
\frac{1}{2\sqrt{\omega}}\sqrt{\tilde{g}^{2}+\tilde{g}^{\prime 2}} 
\label{VII34}
 \end{equation}

Turning now to the fermionic sector, once again we can find the 
spin-connection $\omega_{\mu}^{cd}$. Varying the action with respect to 
$\omega_{\mu}^{cd}$ we obtain an equation similar to Eq. (\ref{C2}) and 
the appropriate modification of Eqs. (\ref{C4})-(\ref{C7}) as the 
solution of it. In the Einstein frame the $\sigma$-contribution
$K_{\mu}^{cd}(\sigma)$ disappears as we discussed at the end of Sec. IVA.  
Contribution of the fermions  selfinteraction to $\omega_{\mu}^{\prime ab}$, 
like 
$\propto\kappa\eta_{ci}V_{d\mu}^{\prime}
\varepsilon^{abcd}\overline{e_{L}^{\prime}}
\gamma^{5}\gamma^{i}e_{L}^{\prime}$ 
\quad ($e_{L}^{\prime}$ is the left electron spinor in the Einstein frame)
which are suppressed by factor $\kappa=16\pi G$ 
in comparison to the first term in Eq. (\ref{VI9}), may be neglected 
if we are interested in particle physics at energies much less than the 
Planck energy scale. 
Therefore neglecting the last term in Eq. (\ref{VI9}) we remain only 
with the Riemannian contribution to the spin-connection.  

Concerning the mass generation of fermions, we have to point out that the 
choice of the last term in Eq. (\ref{VII18}) is related to our intention 
to use the mechanism for the electron mass generation developed in 
Sec.IVC, where we have found out that the exponent $q$ must be equal to
2/3. 
Comparing 
with notations of Sec.IVC we see that after the SSB, 
the factor $C$ in Eq. (\ref{VI18}) has to be identified in the low energy 
theory as 
\begin{equation}
km^{2}=C=\lambda_{e}\left(\frac{\eta}{\sqrt{2}}\right)^{2/3}m^{4/3}
\label{VII35}
 \end{equation}
As a result we get from Eq. (\ref{VI17}):
\begin{equation}
m_{e}=\frac{\lambda_{e}^{3/2}}{\sqrt{6u_{0}\omega}}\eta
\label{VII36}
 \end{equation}

Notice that ratios between masses of all particles of the model (see Eq. 
(\ref{V28}) for the Higgs boson mass,\, Eqs. (\ref{VII33})), (\ref{VII34}) 
for $W^{\pm}_{\mu}$ and $Z_{\mu}$ masses and the electron mass, Eq. 
(\ref{VII36})) are $\omega$-independent. The same is true for the weak angle
\begin{equation}
\sin\theta_{W}=\frac{g^{\prime}}{\overline{g}}
\label{VII37}
 \end{equation}

It is very interesting that a big value of the integration constant 
$\omega$ pushes gauge coupling constants and all masses to small values. In 
addition, the constant $g_{Y}$ of the effective Yukawa coupling 
$g_{Y}\overline{L}^{\prime}e^{\prime}_{R}\varphi$
\begin{equation}
g_{Y}=\frac{\lambda^{3/2}_{e}}{\sqrt{3u_{0}\omega}}
\label{VII38}
 \end{equation}
has an additional factor $u_{0}^{-1/2}$, which can explain (if the gauge 
complex condensate $u_{0}=\frac{y_{0}}{m^{4}}$ is also big) the observed 
further suppression of the effective Yukawa coupling and therefore the 
appropriate suppression of the observed lepton masses. We can think about 
all these effects related to a big values of $\omega$ and $u_{0}$ as a 
{\em "cosmological seesaw mechanism"}, where masses are driven to small 
values due to the appearance of large number in  denominators.

As we pointed out at the beginning of this section, other gauge unified 
theories can be formulated in the same fashion. If  for example we would  
consider also QCD, then the same effect of additional suppression would 
be  obtained for the masses of quarks.  

\section{The true vacuum energy density
as an effect of the SGMF symmetry breaking}

In the previous section we have seen that the presence of the standard 
cosmological
constant term in the original action of the weak NGVE theory does not
change the result of the strong NGVE theory: the vacuum energy density of
the TVS (in the Einstein picture) is zero. We are going to show now that
the appearance of the nonzero vacuum energy density in the TVS (that
is the effective cosmological term) in the Einstein picture  can be the
result of the explicit breaking of the SGMF symmetry (\ref{SG}) or
(\ref{SG1}) by adding the simplest form of a SGMF symmetry breaking term
(\ref{Break}) to the action of the form (\ref{Action12}). Since such term
is invariant
under the LES, it does not contribute to the constraint as one can check
explicitly, and therefore the constraint coincides with Eq. (\ref{CW}).

So, let us consider the model with the action
\begin{equation}
S=\int\Phi
d^{4}x\lbrack
-\frac{1}{\kappa}R(\Gamma, g)+
\frac{1}{2}\varphi_{,\alpha}\varphi^{,\alpha}-V(\varphi)
\rbrack+\int\Lambda\sqrt{-g}d^{4}x-
\gamma\int\frac{\Phi^{2}}{\sqrt{-g}}d^{4}x
\label{BSG}
\end{equation}
Variation with respect to the measure fields $\varphi_{a}$ leads to the
equation
\begin{equation}
A^{\mu}_{a}\partial_{\mu}\lbrack -\frac{1}{\kappa}R(\Gamma, g)
+\frac{1}{2}\varphi_{,\alpha}\varphi^{,\alpha}-V(\varphi)
-2\gamma\frac{\Phi}{\sqrt{-g}}\rbrack=0
\label{BSG1}
\end{equation}

Similar to the cases of  the  strong as well as the weak NGVE theory, it
follows from Eq. (\ref{BSG1}) that if the measure $\Phi\neq 0$,
\begin{equation}
-\frac{1}{\kappa}R(\Gamma, g)
+\frac{1}{2}\varphi_{,\alpha}\varphi^{,\alpha}-V(\varphi)
-2\gamma\chi =M
\label{BSG2}
\end{equation}   

Varying the action Eq. (\ref{BSG}) with respect to $g^{\mu\nu}$ we get
\begin{equation}
-\frac{1}{\kappa}R_{\mu\nu}(\Gamma)
+\frac{1}{2}\varphi_{,\mu}\varphi_{,\nu}
-\lbrack\frac{\Lambda}{2\chi}+\frac{\gamma}{2}\chi\rbrack g_{\mu\nu} =0
\label{BSG4}
\end{equation}

>From Eqs. (\ref{BSG2}) and (\ref{BSG4}) we obtain the constraint which
coincides in the form with the constraint (\ref{CW}) of Sec. IIIB. The
 scalar field $\varphi$ equation has the form of Eq. (\ref{SE}).

In the Einstein frame which still is defined by the conformal
transformation (\ref{Ch3}), after using the constraint (\ref{CW}), we get
the canonical form of equation for the scalar field (\ref{sw1})
with $V^{\prime}_{eff}=\frac{1}{2\Lambda}[M+V(\varphi)]V^{\prime}$  and of
the gravitational equations in the Riemannian space-time with metric
$\overline{g}_{\alpha\beta}$
\begin{equation}
R_{\mu\nu}(\overline{g})-\frac{1}{2}\overline{g}_{\mu\nu}R(\overline{g})=
\frac{\kappa}{2}\lbrace\phi,_{\mu}\phi,_{\nu}-
\frac{1}{2}\overline{g}_{\mu\nu}\phi,_{\alpha}\phi,_{\beta}
\overline{g}^{\alpha\beta}+\lbrack\frac{1}{4\Lambda}(V+M)^{2}
+\gamma\rbrack\overline{g}_{\mu\nu}\rbrace
\label{GB}
\end{equation}

In the TVS $\varphi=\varphi_{0}$ where $V(\varphi_{0})+M=0$ the last term
in Eq.(\ref{GB}) acts as an effective cosmological term.

\section {Model with continuous symmetry related to the NGVE principle
and SSB
without generating a massless scalar field}

It is interesting to see what happens in the above models for the choice
$V=J\phi$, where $J$ is some constant. For simplicity we consider here
this model in the context of the weak NGVE theory. Then the action
(\ref{AW})
is invariant (up to an integral of a total divergence) under the shift
$\phi\rightarrow\phi +const$ which is in fact the symmetry
$V\rightarrow V+const$ related to the NGVE principle. Notice that if we
 consider
the model with complex scalar field $\psi$, where $\phi$ is the phase of
$\psi$, then the symmetry $\phi\rightarrow\phi +const$ would be the
$U(1)$ - symmetry of the classical equations of motion.

The effective potential $V_{eff}(\phi)=
\frac{1}{4\Lambda}[M+V(\phi)]^{2}$ in such a model has the form
\begin{equation}
V_{eff}=\frac{1}{2}m^{2}(\phi -\phi_{0})^{2}
\label{18}
\end{equation}
 where $\phi_{0}=-M/J$
and $m^{2}=J^{2}/2\Lambda$. We see that the symmetry
$\phi\rightarrow\phi +const$ is spontaneously broken and mass
generation
is obtained. However, {\em no massless scalar field results from
the
process of SSB in this case, i.e. Goldstone theorem does not apply
here}. Of course the conditions under which this theorem was proved do not
apply here. In particular, the consideration of a dynamical measure opens
up a totally different way to spontaneously break the $U(1)$ symmetry.

This seems to be a special feature of the NGVE - theory which
allows:
1) To start with linear potential $J\phi$ without destroying the
shift
symmetry $\phi\rightarrow\phi +const$, present in the
$\partial_{\mu}\phi \partial^{\mu}\phi$ piece, due to the coupling to the
dynamical measure (\ref{Fi}). This shift symmetry is now a symmetry of
the
action up to a total divergence.
2) This potential gives rise to an effective potential
$(M+J\phi)^{2}/4\Lambda$. The constant of integration $M$ being
responsible for the SSB.

In the model of Sec.VI for the choice $V=J\phi$, the only difference is
the appearance of the $\lambda$-term in addition to Eq.(\ref{18}).
Similar effect can be obtained also in the strong NGVE - theory
 but
with the
use of 4-index field strength condensate. The possibility of constructing
spontaneously broken $U(1)$ models which do not lead to associated
Goldstone bosons is of course of significant physical relevance. One may
recall for example the famous $U(1)$ problem in QCD \cite{Col}. Also
the possibility of mass generation for axions is of considerable
interest. These
issues will be developed further in elsewhere.
\bigskip
\section{Discussion and Conclusions}

\bigskip
1. {\em Local Einstein Symmetry, Constraint and Degrees of Freedom}. 
In this paper we have seen that the idea to allow the measure to be 
determined dynamically rather than postulating it to be $\sqrt{-g}$ from 
the beginning, has deep consequences. In fact, in the context of the 
first order formalism
the NGVE theory does not have a CCP. This is due
to the fact that the constraint appears which plays a fundamental role in
the theory.
With respect to the scalar field $\chi$ which is built from the dynamical
measure $\Phi$ and the GR measure $\sqrt{-g}$ \quad ($\chi
=\frac{\Phi}{\sqrt{-g}}$),
the theory behaves in three different ways depending on
the structure and symmetries of the action: 

i) If we start from the action (\ref{Action}), that is working in the
framework of the strong NGVE theory, and the Lagrangian density is such
that the constraint is automatically satisfied, then no information about
$\chi$ is obtained and as it turns out, in this case there is a so called
local
Einstein symmetry (\ref{ES1}), (\ref{ES2}) which allows us to set $\Phi$
to whatever we want and in particular $\Phi =\sqrt{-g}$ is possible.

ii) The second possibility exists, as we have seen in Sec. IIC, where
$\chi$ does not enter the constraint, which imposes a condition on other
fields of the theory.

iii) In the context both the strong and the weak NGVE theory, we
presented several models where the dynamical measure $\Phi$ enters in the
constraint (through the field $\chi$) and as a result, it is possible to
solve the field $\chi$ in terms of other fields. It is very important to
note that in such models the constraint describes a connection between
the geometrical object (dynamical measure $\Phi$) and matter fields,
{\em without the Newton constant}. This points out to the fact that the
constraint
affects the physical processes at all region of energy, including low
energy physics. 

Although we allow the dynamical measure $\Phi$ to carry independent
degrees of
freedom (from the variational point of view), in all the listed cases 
after the equations of motions are analyzed, it
turns out that  the
dynamical measure $\Phi$ does
 not introduce new degrees of freedom. Instead, it is responsible for the
rearrangement of interactions in such a way that the CCP problem is
solved when the theory is put (by the appropriate change of variables) in
the Einstein form where both
gravitational constant and all masses are constant. Side by side with
the solution of the CCP problem, the possibility to get
the exact GR form of equations seems to be a very important feature of
the NGVE theory when comparing with usual scalar-tensor theories,
where if one chooses
the Newton constant to be space-time independent (by the use of a change
of variables which involves a conformal transformation), then masses are
space-time dependent and vice versa. The above remarks concern the
examples discussed in this paper but 
this does not mean that the NGVE theory  cannot lead to a non Einstein
behavior (see for example Ref. \cite{G11}). 

Realistic theories are obtained only for the case (iii) as we have seen in
examples (Secs. IID, IIIB, IIIC, V, Appendixes C and D). This has been
achieved in
several ways: in the context of the strong NGVE theory, the introduction
of a 4-index field strength makes it possible to solve $\Phi$ from the
constraint in terms of $g_{\mu\nu}$ and matter fields. In the context of
the weak NGVE theory, the presence of terms in $L_{2}$ which enter the
constraint, makes it possible to solve $\Phi$, again in terms of
$g_{\mu\nu}$ and matter fields.

2.{\em The use of the 4-index field strength in the strong NGVE theory.}    
If a four 
index field strength is introduced in a $4-D$ space - time, it develops a
condensate which turns 
out to be expressed in terms of other fields. As a consequence of this, 
there is 
the possibility to produce the standard particle physics and
gravity dynamics. 
The 
resulting dynamics has then interesting consequences in what concerns to 
the hierarchy problem. As we have seen, all masses and gauge coupling 
constants are driven to small values if the integration constant $\omega$, 
that parameterizes the condensate
strength  (see Eq. (\ref{V21})), is big. In addition to 
this,  masses of fermions are 
driven to small values in comparison with masses of bosons as the gauge 
complex condensate $y_{0}$ becomes big (see Eq. (\ref{VII38})).

The appearance of the parameter $\omega$ in the relation between the 
original coupling constants and the effective ones 
suggests an idea that it may be possible to think in different ways 
concerning renormalization theory. It seems to promise allure prospects 
since the strength of the condensate  specifies a boundary condition or 
state of the Universe.

It is very interesting that a theory designed in order to solve the 
cosmological constant problem tells us about an apparently unrelated
subject, 
like what determines the effective coupling constants and masses of the 
theory. One should recall that the wormhole approach to the 
cosmological constant problem ends up claiming that wormholes determine 
all couplings of the theory also \cite{worm}. 

Other important consequence of the theory described in this paper is 
that one can obtain  scalar field dynamics which allows for an 
inflationary era, the possibility of reheating after scalar field 
oscillations and the setting down to a zero cosmological constant phase 
at the later stages of cosmological evolution, without fine tuning. It is 
interesting that the model not just reproduces all possibilities well 
known in the cosmology of the very early universe solving at the same 
time the cosmological constant problem. In addition to this, the effective 
scalar 
field potential includes integration constants - a feature which makes 
us  hope that by an appropriate choice of those constants, the correct 
density perturbations and reheating could be obtained naturally. Moreover, 
the integration constant $\omega$ enters both in the effective potential 
and in the effective coupling constants and masses. This means that may 
be  a strong correlation between coupling constants and masses of particles 
and some of the cosmological parameters.

Some open problems are of course apparent. For example,  the "persistent 
gauge field condensate scenario" (Sec. IID) has to be understood in a 
deeper way. There a nonspecified function $f$ of a special combination $y$ 
(Eqs. 
(\ref{V1}) or (\ref{VII30}) as examples) of all gauge fields , including 
3-index potential, is 
introduced. For this function we require only the existence of an 
extremum at some point $y=y_{0}>0$ and this is enough to get the
effective action of electro-weak, QCD and other gauge unified models. The 
possible origin of such structure 
as well as the choice of function has to be studied. 

The fact that 
similar type of function appears for example in the QCD effective action as 
the result of  radiative corrections \cite{QCDcondens}, is  encouraging.
In  
such a case no four index field strength is introduced however and the 
effective action is a function of $F_{i\mu\nu}F_{i}^{\mu\nu}$ there. 
Notice that appearance of a four index field strength condensate in the 
framework of the theory developed in this paper, makes the Lorentz 
invariance of the vacuum in QCD not a problem, as opposed to the 
situation where only regular gauge fields are present as the argument of the 
nontrivial 
function, leading to an expectation value of $F_{i\mu\nu}F_{i}^{\mu\nu}$

The four index field strength plays the central role in 
this model. In this connection, one has to recall 
that four index field strength plays a fundamental role in some 
supergravity models, in particular in the $D=11$, $N=1$ supergravity and
in 
the $D=4$, $N=8$ supergravity theories. The possibility of incorporating 
some versions of supergravity into the framework developed in this paper 
seems therefore a subject which could be a potentially fruitful one.    

3. {\em The weak NGVE theory.} The very interesting
effect
we observe in the weak NGVE theory (Sec. III). The addition to the
strong NGVE action (\ref{Action}) an explicit cosmological
constant term of the canonical GR form $\int\Lambda\sqrt{-g}d^{4}x$ not
only
does not prevent from the resolution of the CCP in a way similar to what
was done in the context of the strong NGVE theory, but it  also delivers
us from the need to incorporate the  four index field strength in order to 
get a nontrivial resulting dynamics. Moreover, such an explicit
cosmological constant term results in  the effective potential
of the form (\ref{VW})
which for any analytic original potential $V(\varphi)$ and for $\Lambda
>0$ allows for the scalar field $\varphi$ to reach in its dynamical
evolution the TVS with zero
energy density while no fine tuning is needed.
Of course, the cosmological consequences of the weak NGVE theory developed
in Sec. III
are similar
to those of the strong NGVE theory with the four - index field strength.
However, the absence of the integration constant $\omega$ \quad (Eq.
\ref{V21}) does not allow a strong correlation between coupling constants
and masses of particles
and some of the cosmological parameters as it is expected in the strong
NGVE theory with the four - index field strength.

4. {\em The TVS with nonzero energy density and SGMF symmetry breaking.} 
In order to generate a nonzero energy density in the TVS one has to break
the infinite dimensional SGMF symmetry (\ref{SG}) or (\ref{SG1}) in the
original action. In the simplest model (Sec. VI) this is achieved by
adding
a SGMF symmetry breaking term (\ref{Break}). If this violation of the
SGMF symmetry appears
as a result of quantum
corrections, it represents then the appearance of an anomaly of the  SGMF
symmetry (\ref{SG}) or (\ref{SG1}). We then
have a reason for the smallness of such terms if not of their absolute
vanishing ( in the case of exact symmetry). This resembles the situation
of quark masses and chiral invariance (CI). In this case, as it is well
known, CI forbids a quark mass. If a quark mass nevertheless appears, CI
ensures that quark masses remain small even after the consideration of
quantum corrections. In a similar way if a small SGMF symmetry
breaking term appears, hopefully, it will not be renormalized into a large
contribution after quantum effects are considered.

5. {\em Remarks concerning the "energy - volume complementarity" and
on quantization}
It is interesting that the mechanism that is responsible for canceling
the vacuum energy, which requires $\chi\equiv\Phi
/\sqrt{-g}\rightarrow\infty$
as we approach the TVS, can give also some dramatic effects when we
consider high energy - densities states. To see this, let us take as an
example the model of  Sec. IIIB where Eq. (\ref{CW}) gives $\chi\propto
(V+M)^{-1}$ (consideration of any of the other examples of Sec. IID,
and Appendixes C, D lead to the same qualitative conclusions).

As $V+M\rightarrow 0$, $\chi\rightarrow\infty$ as mentioned before and
this corresponds to the TVS with zero energy density. In the case of high
energy density, that is for example allowing formally
$V+M\rightarrow\infty$, we obtain that $\chi\equiv\Phi
/\sqrt{-g}\rightarrow 0$, in the context of this classical model. Since
$\Phi d^{4}x$ has the interpretation of a volume element in the NGVE
theory as we have
discussed in Introduction and $\sqrt{-g}d^{4}x$ is a standard volume
element of GR, we see that where energy densities go to infinity, the
volume of the NGVE theory becomes zero relative to the GR volume.

If we were study the behavior of $\chi\equiv\Phi/\sqrt{-g}$
near a high concentration of energy of a point-like particle for example,
the same effect is generally obtained, that is the NGVE - volume element
goes to zero at short distances from such particle.

This type of effect
is generic to a wide class of models mentioned before and in addition to
other cases including also models of extended objects\footnote{Notice
that when $\Phi\neq 0$ then $L=const.$ \quad (or $L_{1}=const.$) and the
symmetry (\ref{SG}) (or (\ref{SG1})) represents just a trivial shift of
$\varphi_{a}$ by
a
constant. This shift becomes nontrivial if we allow for points where
$\Phi =0$ and therefore $L$ or $L_{1}$ can vary in such "defect" points.
Such
defects can
be naturally associated with extended objects and the nontrivial symmetry
(\ref{SG}) or (\ref{SG1}) appears then like invariance under
reparametrization of the
extended object coordinates $\varphi_{a}$.}
which provide
regions with singular energy density where the constraint generally tells
us $\chi\rightarrow 0$ as energy density approaches infinity. We will call
this effect the "energy - volume complementarity" which means that in a
certain sense space - time tries "to run away" from regions of extremely
high energy density.

This suggests that in the quantum theory, working with $\Phi$,
$g_{\mu\nu}$ (and not with the effective Einstein variable
$\overline{g}_{\alpha\beta}=\chi g_{\alpha\beta}$, undefined for $\chi
=0$), an interesting effect takes place: where ultra high energy quantum
fluctuations occur, space - time volume approaches zero, thus damping in
the sense of regularization of those very high excitations.

All this indicates that the correct variables in the high energy regime
are the original metric $g_{\alpha\beta}$, the connections
$\Gamma^{\lambda}_{\mu\nu}$, the measure fields $\varphi_{a}$ (and the
matter fields of course). 

In the states which allow {\em regions with} $\Phi =0$, which, as we have
seen, enter in the
complete set of solutions, only $g_{\alpha\beta}$ (and not
$\overline{g}_{\alpha\beta}$) can be used as the dynamical variable.
There {\em and only there} the infinite dimensional SGMF symmetry
(\ref{SG}) or (\ref{SG1}) becomes nontrivial since $L$ or $L_{1}$ are
not
necessarily constants: if $\Phi =0$, then the equation of the type of Eq.
(\ref{FEM}) does not imply $\partial_{\mu}L=0$ or $\partial_{\mu}L_{1}=0$
 (see Eqs. (\ref{L1}) and (\ref{Action12}) for notations). These states
therefore play a crucial role for restrictions on the quantum dynamics
since much of the structure of the theory is restricted by the existence
of the SGMF symmetries (\ref{SG}) or (\ref{SG1}).

Summarizing the previous discussion we conclude that states for which
$\Phi =0$ is achieved somewhere, correspond to states in the extreme
energy density region and the SGMF symmetries (\ref{SG}) or (\ref{SG1})
act
non
trivially. In the quantum theory, we expect therefore that the requirement
that such symmetries be preserved by, for example, certain regularizations
of the theory, will ensure the preservation of the structure of the theory
which is responsible for the solution of the CCP.

Notice finally that in the context of the weak NGVE principle, the
infinite dimensional SGMF symmetry still exists in the form specified in
Eq. (\ref{SG1}). This suggests that the weak NGVE model can be a
realization of the effective action that takes into account quantum
corrections.

6. {\em A complete group-theoretical and/or algebraic study} of the
infinite
dimensional symmetries of the theory has not been carried out yet and
this should be an interesting exercise. One should notice that in
addition to (\ref{SG}) or (\ref{SG1}) there is the infinite dimensional
symmetry of
volume preserving internal diffeomorphisms (VPD)
$\varphi^{\prime}_{a}=\varphi^{\prime}_{a}(\varphi_{b})$ such that
$\Phi^{\prime}=\Phi$. Such transformations applied after a transformation
of the form (\ref{SG}) or (\ref{SG1}) lead to something new. The full
symmetry group of
transformation contains element that are not in SGMF and not in VPD which
are only subgroups of the yet unknown full group of internal symmetries
of
the measure scalars $\varphi_{a}$.

7. {\em In our treatment of the symmetry $L\rightarrow L+const.$}
we have
ignored possible topological effects, since $\int\Phi d^{4}x$ being a
total divergence can be responsible however for topological effects,
similar to the well known $\Theta$-term in QCD.

As we have seen in Sec. VII, such symmetry can be exploited to
construct (through the introduction of a linear potential $V=J\phi$
coupled to the measure $\Phi$) a theory that is globally $U(1)$-invariant
with SSB and yet without appearance of a Goldstone boson. Possible
applications to axion mass generation, etc has to be explored.

 \bigskip
\section{Acknowledgments}
We thank Profs. J. Bekenstein, R. Brustein, A. Davidson, A. Dolgov, S.
Elitzur, A.
Guth, F. Hehl, P. Mannheim,
Y. Ne'eman, E. Nissimov, S. Pacheva and R. Zalaletdinov for 
useful and encouraging discussions on the subjects of this paper.

\bigskip

\appendix

\section{Connection to other developments concerning volume preserving
diffeomorphisms}

It is interesting to compare the composite density (\ref{Fi}) and the form
of the equations of motion obtained from the variation of the action
(\ref{Action}), with other composite structures and their associated
equations of motion. 

The simplest instance of a composite structure is
obtained by considering two scalar fields $\varphi_{i}$ ($i=1,2$) and
defining the composite field strength \cite{GNP}
\begin{equation}
F_{\mu\nu}=\epsilon_{ij}\partial_{\mu}\varphi_{i}\partial_{\nu}\varphi_{j}
        \label{GNP1}
\end{equation} 
where $\epsilon_{ij}$ is the two dimensional antisymmetric symbol with
$\epsilon_{12}=1=-\epsilon_{21}$, $\epsilon_{11}=\epsilon_{22}=0$. The
form (\ref{GNP1}) derives from a vector potential 
\begin{equation}
F_{\mu\nu}=\partial_{\mu}A_{\nu}-\partial_{\nu}A_{\mu}
        \label{GNP2}
\end{equation}
 without
need of imposing an additional condition, for example with
\begin{equation}
A_{\mu}=\frac{1}{2}\epsilon_{ij}\varphi_{i}\partial_{\mu}\varphi_{j},
        \label{GNP3}
\end{equation}
all without reference to the dimensionality of space - time.

The form (\ref{GNP1}) is associated with an area, that is, a two
dimensional volume. In fact, $F_{\mu\nu}$ given by (\ref{GNP1}) is
invariant under the area preserving diffeomorphisms, i. e. internal
transformations in the $\varphi_{i}$ space
\begin{equation}
\varphi_{i}\rightarrow\varphi_{i}^{\prime}=\varphi_{i}^{\prime}
(\varphi_{j})
        \label{GNP4}
\end{equation}
which satisfy the area preserving condition
\begin{equation}
\epsilon_{ij}\frac{\partial \varphi_{i}^{\prime}}{\partial \varphi_{k}}
             \frac{\partial \varphi_{j}^{\prime}}{\partial \varphi_{l}} 
 =\epsilon_{kl}
        \label{GNP5}
\end{equation}

It is not a surprise then that under such transformation $A_{\mu}$
is transformed by the addition of a gradient \cite{GNP}
\begin{equation}
A_{\mu}\rightarrow A_{\mu}+\partial_{\mu}\Xi
        \label{GNP6}
\end{equation}
where $\Xi$ is determined by the area preserving transformation
(\ref{GNP4}).

One can realize then many of the concepts standard in gauge theories
(including minimal coupling to charged matter),
where now the gauge symmetry is replaced by the area preserving
diffeomorphisms. The number of scalars $\varphi_{i}$ can be enlarged from
2 to $2n$
($n=1,2,...$) and then
$F_{\mu\nu}=\omega_{ab}\partial_{\mu}\varphi_{a}\partial_{\nu}\varphi_{b}$
where
$\omega_{ab}$ has to be a $2n\times 2n$ block
diagonal antisymmetric matrix, being equal to $\epsilon_{ab}$ in  each
block \cite{GNP}.
An interesting case in $4-D$ is when we deal with 4 scalars $\varphi_{a}$
($a=1,2,3,4$) and
\begin{displaymath}
\omega_{ab}=
\left( \begin{array}{cccc}
 0 & 1 & 0 & 0 \\
-1 & 0 & 0 & 0 \\
 0 & 0 & 0 & 1 \\
 0 & 0 &-1 & 0  
\end{array} \right)
\label{GNP7}
\end{displaymath}
as we will see.

If we take the Lagrangian density in a flat four dimensional space - time
as
\begin{equation}
L=-\frac{1}{4}F_{\mu\nu}(\varphi_{a})F^{\mu\nu}(\varphi_{a})
\label{GNP8}
\end{equation}
and the action $S=\int Ld^{4}x$, then we get from the variation with
respect to $\varphi_{a}$ the equations
\begin{equation}
\partial_{\alpha}\varphi_{a}(\partial_{\beta}F^{\beta\alpha})=0
\label{GNP9}
\end{equation}

For the purpose of comparing with the NGVE theory, let us consider the
situation with $n=2$. In this case $\det (\partial_{\alpha}\varphi_{a})$
is well defined in $4-D$ space - time. In fact the condition that
(\ref{GNP9}) is equivalent to Maxwell's equation is that $\Phi\neq 0$ with
$\Phi$ being defined by Eq. (\ref{Fi}). It has been shown \cite{GNP} that
solutions where $\Phi =0$ exist and they can produce effects like mass
generation etc.. As we are pointing out in the present paper, the $\Phi
=0$ solutions in the NGVE theory are expected to play an important role
and the fact that similar type of solution have been found in a simpler
theory, gives us hope that future research in this subject has good
possibilities.

Composite gauge fields of the type discussed here have also been found
useful applications in the context of brane theory \cite{Castro}. Further
discussion on interrelations between the NGVE theories, composite gauge
theories and other developments can be found in \cite{G2}. 

\bigskip

\section{Metric - Affine formalism in the NGVE theory and $\lambda$-symmetry}

In Sec. IIB we have shown that in the NGVE theory there is a 
contribution of the $\chi$-field to the connection (see Eqs. (\ref{GAM2}) 
and (\ref{S2})). This contribution is defined up to the 
$\lambda$-symmetry transformation (\ref{Gamal}). By using this symmetry, 
in Sec.IIB we have chosen the gauge where the antisymmetric part of the 
connection (that is a $\chi$-contribution into the torsion) disappears.

It is interesting to see what is the geometrical meaning of the 
$\lambda$-gauge dependent contribution to the connection (\ref{GAM2}), 
(\ref{S2}). For this we calculate the covariant derivative of the 
metric tensor $g_{\mu\nu}$ with the connection defined by (\ref{GAM2})
and (\ref{S2}) and we get
\begin{equation}
g_{\mu\nu ;\alpha}=-2g_{\mu\nu}\lambda_{,\alpha}\equiv N_{\mu\nu\alpha}.
        \label{A1}
\end{equation}
This means that the $\lambda$ dependent contribution to the connection
is responsible for the appearance of the nonmetricity tensor \cite{nonmetr}.

With the choice $\lambda = \frac{1}{2}\sigma$, in Sec. IIB, we have 
eliminated the $\chi$-contribution into the torsion keeping the 
nonmetricity tensor $N_{\mu\nu\alpha}$ which in this 
"$\sigma$-torsionless" gauge is equal to \quad 
$-g_{\mu\nu}\sigma_{,\alpha}$. However we have the freedom to choose for 
example the "$\sigma$- metric - compatible" gauge where the nonmetricity 
disappears: $\lambda =constant$. In such a case, the torsion is not 
eliminated from the connection (\ref{GAM2}), (\ref{S2}).

 Notice that these peculiarities of the $\lambda$-symmetry concerning the 
possibility of eliminating the $\chi$-contribution to the torsion or, 
alternatively, of eliminating the $\chi$-contribution to the nonmetricity 
appear to be a very interesting feature of the NGVE theory in the metric 
- affine formalism. This feature results from the basic assumption that 
not only metric and connection are independent dynamical variables (as it 
is in the case of the Metric - Affine theory), but also the measure degrees 
of freedom are independent variables when varying the action. 

\bigskip
\section{Four index field strength condensate as an universal governor. 
\protect\\
Simple models.}
\bigskip

\subsection{Scalar field + four index field strength}
 \bigskip

As it follows from our analysis in  Sec.IIC, a model with only a
scalar field although solves the cosmological constant problem, it
cannot give a nontrivial dynamics.
We study here the simplest model which in addition to  a single scalar
field
with a nontrivial potential includes also an additional degree of freedom
described by a three-index potential $A_{\beta\mu\nu}$  as in the Lagrangian
density
\begin{equation}
L=-\frac{1}{\kappa}R(\Gamma,g)
+\frac{1}{2}\varphi,_{\alpha}\varphi^{,\alpha}-V(\varphi)+
\frac{1}{4!}F_{\alpha\beta\mu\nu}F^{\alpha\beta\mu\nu}.
 \label{IV1}
\end{equation}

Here 
\begin{equation}
F_{\alpha\beta\mu\nu}\equiv\partial_{[\alpha}A_{\beta\mu\nu ]}
\label{IV2}
\end{equation}
 is
the field strength which is invariant under the gauge transformation
\begin{equation}
A_{\beta\mu\nu}\rightarrow A_{\beta\mu\nu}+\partial_{[\beta}f_{\mu\nu]}
\label{IV3}
\end{equation}

In ordinary 4-dimensional GR, the
$F_{\alpha\beta\mu\nu}F^{\alpha\beta\mu\nu}$ term gives rise
to a  cosmological constant depending on an integration constant
\cite{FF1},\cite{FF2}. In our case,
due to the constraint (\ref{II3}), the degrees of freedom contained in
$F_{\alpha\beta\mu\nu}$ and those of the scalar field $\varphi$ will be
intimately correlated. The sign in front of the
$F_{\alpha\beta\mu\nu}F^{\alpha\beta\mu\nu}$ term is chosen so that in
this model the
resulting expression for the energy density of the scalar field $\varphi$ is
a positive definite one for any possible
space-time dependence of $\varphi$ in the Einstein picture". Notice also
that two last terms in the action with the Lagrangian (\ref{IV1}) break
explicitly the LES.

The gravitational equations (\ref{II2}) take now the form
\begin{equation}
-\frac{1}{\kappa}R_{\mu\nu}(\Gamma)
+\frac{1}{2}\varphi,_{\mu}\varphi,_{\nu}+
\frac{1}{6}F_{\mu\alpha\beta\gamma}F_{\nu}^{\alpha\beta\gamma}=0.
 \label{IV4}
 \end{equation}
Notice that the scalar field potential $V(\varphi)$ does not appear
explicitly in Eqs. (\ref{IV4}). However, the constraint (\ref{II3}), which
takes now  the form
\begin{equation}
V(\varphi)+M=
-\frac{1}{8}F_{\alpha\beta\mu\nu}F^{\alpha\beta\mu\nu},
 \label{IV5}
 \end{equation}
allows us to express the last term in (\ref{IV4}) in terms of the
potential $V(\varphi)$ (using the fact that
$F^{\alpha\beta\mu\nu}\propto\varepsilon^{\alpha\beta\mu\nu}$ in
4-dimensional space-time).

Varying the action with respect to $A_{\nu\alpha\beta}$, we get the equation
\begin{equation}
\partial_{\mu}(\Phi F^{\mu\nu\alpha\beta})=0
\label{IV6}
 \end{equation}
 Its general solution has
a form
\begin{equation}
F^{\alpha\beta\mu\nu}=
\frac{\lambda}{\Phi}\varepsilon^{\alpha\beta\mu\nu}
\equiv\frac{\lambda}{\chi\sqrt{-g}}\varepsilon^{\alpha\beta\mu\nu} ,
 \label{IV7}
 \end{equation}
where $\lambda$ is an integration constant. Then
$F_{\alpha\beta\mu\nu}F^{\alpha\beta\mu\nu}=-\lambda^{2}4!/\chi^{2}$ 
is not a constant now as
opposed to the GR case \cite{FF1},\cite{FF2}  and
therefore
\begin{equation}
V(\varphi)+M=3\lambda^{2}/\chi^{2}
 \label{IV8}
 \end{equation}
and
\begin{equation}
F_{\mu\alpha\beta\gamma}F_{\nu}^{\alpha\beta\gamma}=
-(6\lambda^{2}/\chi^{2})g_{\mu\nu}=-2[V(\varphi)+M]g_{\mu\nu}
\label{IV9}
 \end{equation}
This shows how
the potential $V(\varphi)$ appears in Eq. (\ref{IV4}), spontaneously 
violating
the symmetry of the action $V(\varphi)\rightarrow V(\varphi)+constant$, which
now corresponds to a redefinition of the integration constant $M$.

The equation of motion of the scalar field 
$\varphi$ is
\begin{equation}
(-g)^{-1/2}\partial_{\mu}(\sqrt{-g}g^{\mu\nu}\partial_{\nu}\varphi)
+\sigma,_{\mu}\varphi ^{,\mu}+V^{\prime}(\varphi)=0,
\label{IV10}
 \end{equation}
 where
$V^{\prime}\equiv\frac{dV}{d\varphi}$.

The derivatives of the field
$\sigma$ enter both in the gravitational Eqs. (\ref{IV4})
(through the connection) and in the scalar field equation (\ref{IV10}). In 
order to
see easily the physical content of this model, we have to perform a
{\em conformal transformation} 
\begin{equation}
\overline{g}_{\mu\nu}(x)=\chi
g_{\mu\nu}(x); \qquad \varphi\rightarrow\varphi
\label{IV11}
 \end{equation}
 to obtain an Einstein picture.
Notice that now this transformation is not a symmetry and indeed
changes the form of equations.  In
this new frame, the gravitational equations become
\begin{equation}
G_{\mu\nu}(\overline{g}_{\alpha\beta})=\frac{\kappa}{2}T_{\mu\nu}^{eff}(\varphi)
 \label{IV12}
 \end{equation}
where
\begin{equation}
G_{\mu\nu}(\overline{g}_{\alpha\beta})=
R_{\mu\nu}(\overline{g}_{\alpha\beta})-
\frac{1}{2}\overline{g}_{\mu\nu}R(\overline{g}_{\alpha\beta})
 \label{IV13}
 \end{equation}
is the Einstein tensor in the Riemannian space-time with metric 
$\overline{g}_{\mu\nu}$, and the source is the minimally coupled scalar 
field $\varphi$
 \begin{equation}
T_{\mu\nu}^{eff}(\varphi)=
\varphi,_{\mu}\varphi,_{\nu}-
\frac{1}{2}\overline{g}_{\mu\nu}\varphi,_{\alpha}\varphi,^{\alpha}
+\overline{g}_{\mu\nu}V_{eff}(\varphi)
 \label{IV14}
 \end{equation}
with the new effective potential
\begin{equation}
V_{eff}=\frac{2}{3}\chi^{-1}(V+M)
 \label{IV15}
 \end{equation}

The scalar field equation (\ref{IV10}) in the Einstein picture takes a form
\begin{equation}
\frac{1}{\sqrt{-\overline{g}}}
\partial_{\mu}(\sqrt{-\overline{g}}\ \overline{g}^{\mu\nu}
\partial_{\nu}\varphi)+
\chi^{-1}V^{\prime}(\varphi)=0.
 \label{IV16}
 \end{equation}

For the possible expression for $\chi^{-1}$ we have from Eq. (\ref{IV8}) 
\begin{equation}
\frac{1}{\chi}=\pm\frac{1}{\lambda\sqrt{3}}(V+M)^{1/2}
 \label{IV17}
 \end{equation}

Independently of the sign chosen for $\chi$ in Eq. (\ref{IV17}), the 
gravitational and scalar field equations have the 
same physical content expressed in different space-time signatures. In 
what follows we simply take the + sign in (\ref{IV17}). Therefore, the 
effective scalar field potential (\ref{IV15}) has the form
\begin{equation}
V_{eff}(\varphi)=\frac{2}{\lambda 3\sqrt{3}}(V+M)^{3/2}
 \label{IV18}
 \end{equation}
and the scalar field Eq. (\ref{IV16}) becomes a conventional general 
relativistic scalar field equation with the potential $V_{eff}(\varphi)$:
\begin{equation}
\frac{1}{\sqrt{-\overline{g}}}
\partial_{\mu}(\sqrt{-\overline{g}}\quad\overline{g}^{\mu\nu}
\partial_{\nu}\varphi)+
V_{eff}^{\prime}(\varphi)=0.
 \label{IV19}
 \end{equation}

We see that in the Einstein picture, for {\em any analytic} $V(\varphi)$, \
$V_{eff}(\varphi)$ has an
extremum, that is $V^{\prime}_{eff}=0$, {\em either} when $V^{\prime}=0$
{\em or\/} $V+M=0$. The extremum when $V+M=0$ corresponds to an absolute
minimum (since $V_{eff}(\varphi)$ is non negative) and therefore it is {\em a
vacuum with
zero effective cosmological constant\/}. It should be emphasized that all
what is required is that $V+M$ reaches zero at {\em some\/} point
$\varphi_{0}$ but $V^{\prime}$ at this point does not  need to vanish.
Therefore {\em no fine tuning\/} in the usual sense, of adjusting a
minimum of  a potential to coincide with the point where this
potential itself vanishes, is required. And the situation is even more
favorable since even if $V+M$ happens not to touch  zero for any value of
$\varphi$, we  always have an infinite set of other values of the 
integration constant $M$ where this will happen.
However, this model is not as satisfactory as those discussed in the text
since the second derivative of $V_{eff}$ is singular in its absolute
minimum.  

\bigskip
\subsection{Gauge fields and the Higgs mechanism in the NGVE theory}

\bigskip

Let us consider now a model including gravity, four index field strength 
$F_{\alpha\beta\mu\nu}$, a 
gauge field $\tilde{A}_{\mu}$ and a complex scalar field $\phi$ minimally 
coupled to the gauge field with the action
\begin{eqnarray}
S=\int\Phi d^{4}x[-\frac{1}{\kappa}R(\Gamma,g)
+\frac{1}{4!}F_{\alpha\beta\mu\nu}F^{\alpha\beta\mu\nu}
+\frac{1}{m^{4}}(\tilde{F}_{\mu\nu}\tilde{F}^{\mu\nu})^{2}
\nonumber\\
+g^{\mu\nu}(\partial_{\mu}-i\tilde{e}\tilde{A}_{\mu})\phi
           (\partial_{\nu}+i\tilde{e}\tilde{A}_{\nu})\phi^\ast
-V(|\phi|)]
 \label{IV23}
\end{eqnarray}
where $\tilde{F}_{\mu\nu}\equiv\partial_{\mu}\tilde{A}_{\nu}-
\partial_{\nu}\tilde{A}_{\mu}$.

Notice that the kinetic term of the gauge field $\tilde{A}_{\mu}$ is 
chosen in an unusual way where an additional parameter $m$ with 
dimensionality of mass is introduced to provide the canonical 
dimensionality for the gauge field $\tilde{A}_{\mu}$. The reason for such 
a choice of the kinetic term of the gauge field is to achieve for it the 
same degree of homogeneity in $g^{\mu\nu}$ in the Lagrangian density as we 
have for the four index field strength $F_{\alpha\beta\mu\nu}$. As we 
will see, for such a choice, after solving the constraint we obtain the 
standard effective low energy physics. For example, in the absence of 
other interactions, the gauge field equations possess conformal 
invariance or, what is the same, they have the standard Maxwell form.

By making use the gauge invariance we choose the unitary gauge (where 
$Im\,\phi (x)=0$) and then the Lagrangian density takes the form
\begin{eqnarray}
L=-\frac{1}{\kappa}R(\Gamma,g)
+\frac{1}{4!}F_{\alpha\beta\mu\nu}F^{\alpha\beta\mu\nu}
+\frac{1}{m^{4}}(\tilde{F}_{\mu\nu}\tilde{F}^{\mu\nu})^{2}+
\nonumber\\
\frac{1}{2}g^{\mu\nu}\varphi,_{\mu}\varphi,_{\nu}-
V(\varphi)+
\frac{1}{2}\tilde{e}^{2}\varphi^{2}g^{\mu\nu}\tilde{A}_{\mu}\tilde{A}_{\nu})
 \label{IV24}
\end{eqnarray}
where we have defined $|\phi| =\frac{1}{\sqrt{2}}\varphi$.

The constraint (\ref{II3}) corresponding to the Lagrangian density 
(\ref{IV24}) is
\begin{equation}
\frac{1}{8}F_{\alpha\beta\mu\nu}F^{\alpha\beta\mu\nu}
+\frac{3}{m^{2}}(\tilde{F}_{\mu\nu}\tilde{F}^{\mu\nu})^{2}+
V(\varphi)+M=0
 \label{IV25}
\end{equation}
Similar to the first and the fourth terms, the last term 
in Eq. (\ref{IV24}) does not contribute to the constraint since it is 
homogeneous of degree one in $g^{\mu\nu}$.

The equation for $F_{\alpha\beta\mu\nu}$ and its solution are still the 
same as in Eqs. (\ref{IV6}), (\ref{IV7}) which bring constraint 
(\ref{IV25}) to the following equation:
\begin{equation}
\frac{\omega^{2}m^{4}}{\chi^{2}}=
\frac{1}{m^{4}}(\tilde{F}_{\mu\nu}\tilde{F}^{\mu\nu})^{2}+
\frac{1}{3}(V(\varphi)+M)
 \label{IV26}
\end{equation}
where we have defined $\lambda$ in terms of the mass parameter $m^{2}$ 
\begin{equation}
\lambda=\omega m^{2},
 \label{IV27}
\end{equation}
$\omega$ being a dimensionless constant.

Varying the action with respect to $\tilde{A}_{\mu}$ we get 
\begin{equation}
\frac{1}{\sqrt{-g}}\partial_{\mu}[\sqrt{-g}\chi 
(\tilde{F}_{\alpha\beta}\tilde{F}^{\alpha\beta})g^{\mu\gamma}g^{\nu\delta}
\tilde{F}_{\gamma\delta}]
-\frac{\tilde{e}^{2}m^{4}}{8}\varphi^{2}\chi g^{\alpha\nu}\tilde{A}_{\alpha}
=0
 \label{IV28}
\end{equation}

Looking at gauge field fluctuations around the true vacuum 
$\varphi=\varphi_{0}$ where $V(\varphi_{0})+M=0$ and 
$V^{\prime}_{eff}(\varphi_{0})=0$ 
(and ignoring the scalar field fluctuations around the true vacuum
$\varphi_{0}$,  see the previous 
subsection), we get from  (\ref{IV26})
\begin{equation}
\chi (\tilde{F}_{\mu\nu}\tilde{F}^{\mu\nu})=
\pm\omega m^{4}
 \label{IV29}
\end{equation}
First of all notice that in the case where is no coupling to the scalar 
field ($\tilde{e}=0$), Eq. (\ref{IV28}) after making use Eq. (\ref{IV29}) 
becomes the Maxwell's equations 
\begin{equation}
\frac{1}{\sqrt{-g}}\partial_{\mu}({\sqrt{-g}}
g^{\mu\gamma}g^{\nu\delta}
\tilde{F}_{\gamma\delta})
=0
 \label{IV30}
\end{equation}
which are indeed conformally invariant.

In the presence of interactions with scalar field ($\tilde{e}\not=0$), 
Eqs. (\ref{IV28}) and (\ref{IV29}) lead to the singularity in the second 
term when $\tilde{F}_{\alpha\beta}\tilde{F}^{\alpha\beta}=0$.

Notice that the $\chi$-field becomes divergent as $\varphi$ 
approaches the absolute minimum $\varphi_{0}$. Therefore it is not a 
surprise that a singularity also occurs in the case where 
$\tilde{F}_{\alpha\beta}\tilde{F}^{\alpha\beta}=0$, that is when electric 
dominated field evolves into magnetic dominated one (or vice versa). 
Here we will see 
that this singularity is eliminated by the  conformal transformation 
(\ref{IV11}) to the Einstein frame.

In fact, performing the conformal transformation (\ref{IV11}) and taking 
into account the constraint (\ref{IV29}) we obtain for the gauge field 
equation in the Einstein frame
\begin{equation}
\frac{1}{\sqrt{-\overline{g}}}\partial_{\mu}[\pm\sqrt{-\overline{g}}
\enspace\overline{g}^{\mu\gamma}\overline{g}^{\nu\delta}
\tilde{F}_{\gamma\delta}]
-\frac{\tilde{e}^{2}}{8\omega}\varphi^{2}\overline{g}^{\alpha\nu}
\tilde{A}_{\alpha}
=0
 \label{IV31}
\end{equation}

Notice that we have not taken the sign $\pm$ in Eq. (\ref{IV31}) outside 
the derivative operator. This makes it apparent that if we pick one of 
the two branches displayed in Eq. (\ref{IV31}), it cannot evolve 
continuously to the another branch. \footnote{We have to point out that
the models discussed in the body of the paper do not suffer this
ambiguity.} Therefore the requirement of the 
analyticity of the resulting equations exclude for example the alternative
$\chi |\tilde{F}_{\mu\nu}\tilde{F}^{\mu\nu}|=
\omega m^{4}$ instead of (\ref{IV29}) as a solution of the constraint 
(\ref{IV26}) at the point $\varphi=\varphi_{0}$.

When choosing one of the branches in Eq. (\ref{IV29}) we note that the 
conformal transformation (\ref{IV11}) changes the relative signatures of 
the original metric $g_{\mu\nu}$ and the metric $\overline{g}_{\mu\nu}$ in 
the Einstein frame when the gauge field evolves from electric dominated 
to magnetic dominated (and vice versa).

Eq. 
(\ref{IV31}) shows that in the Einstein frame there is no singularity 
when $\chi^{-1}\rightarrow\pm 0$. Therefore, changes of the signature of 
the metric take place only in the original frame and not in the Einstein 
frame.

For the choice of the signature $(+---)$ in the Einstein frame we have to 
choose the branch $(-)$ in Eq.
(\ref{IV31}) in order to avoid tachyonic behavior. After the change of 
notations
\begin{equation}
e=\frac{\tilde{e}}{2\sqrt{2\omega}};\qquad 
A_{\mu}=2\sqrt{2\omega}\tilde{A}_{\mu}; 
 \label{IV32}
\end{equation} 
we get the canonical form of equations for the vector field 
\begin{equation}
\frac{1}{\sqrt{-\overline{g}}}\partial_{\mu}(\sqrt{-\overline{g}}
\enspace\overline{g}^{\mu\gamma}\overline{g}^{\nu\delta}
F_{\gamma\delta})
+m_{A}^{2}A^{\nu}
=0
 \label{IV33}
\end{equation}
 where
\begin{equation}
m_{A}^{2}=e^{2}\varphi_{0}^{2}
 \label{IV34}
\end{equation}
is the mass of the vector boson $A_{\mu}$ which is generated by the 
spontaneous symmetry breaking (SSB) of the gauge invariance when the scalar 
field $\phi$ is located at the absolute minimum $|\phi|=
\frac{1}{\sqrt{2}}\varphi =\frac{1}{\sqrt{2}}\varphi_{0}$ of the effevtive 
potential (\ref{IV18}). We have used the notations
\begin{equation}
A^{\mu}=\overline{g}^{\mu\nu}A_{\nu}; \ F_{\mu\nu}=\partial_{\mu}A_{\nu}-
\partial_{\nu}A_{\mu}; \ 
F^{\mu\nu}=\overline{g}^{\mu\alpha}\overline{g}^{\nu\beta}F_{\alpha\beta}
 \label{IV35}
\end{equation} 

The appropriate gravitational equations at the absolute minimum 
$\varphi_{0}$ in the Einstein frame, when the 
source is the vector field $A_{\mu}$ takes the standard form
\begin{equation}
G_{\mu\nu}(\overline{g}_{\alpha\beta})=\frac{\kappa}{2}T_{\mu\nu}
(A_{\alpha})
 \label{IV36}
\end{equation}
where
\begin{equation}
T_{\mu\nu}(A_{\alpha})=
\frac{1}{4}\overline{g}_{\mu\nu}F_{\alpha\beta}F^{\alpha\beta}
-F_{\mu\alpha}F_{\nu\beta}\overline{g}^{\alpha\beta}+
\frac{1}{2}m_{A}^{2}(A_{\mu}A_{\nu}-
\frac{1}{2}\overline{g}_{\mu\nu}A_{\alpha}A^{\alpha})
 \label{IV37}
\end{equation} 

For the scalar field equation in the Einstein frame we obtain 
\begin{equation}
\frac{1}{\sqrt{-\overline{g}}}\partial_{\mu}(\sqrt{-\overline{g}}
\enspace\overline{g}^{\mu\nu}\partial_{\nu}\varphi)-
e^{2}A_{\alpha}A^{\alpha}\varphi+V^{\prime}_{eff}=0
 \label{IV38}
\end{equation}
with $V_{eff}$ given by Eq. (\ref{IV18}).

As we see, the rescaling (\ref{IV32}) provides the canonical 
normalization of the gauge field so to reproduce the standard form of the 
energy-momentum tensor (\ref{IV37}) and standard interaction of the 
vector field $A_{\mu}$ to the scalar field after SSB. A very interesting 
feature of the theory in the Einstein picture is the fact that the gauge 
coupling constant $e$  depends on the integration constant $\omega$ which 
appears in the solution for the four index field condensate (\ref{IV7}) 
(see also the definition (\ref{IV27}).

Finally we have to notice that it is possible to improve the model 
discussed in this Sec.IV (see for example \cite{GK4}) in such a way that 
the mass of the scalar field becomes finite (compare with discussion at 
the end of Subsec.IVA). Then, however, at the very high energies where 
the scalar field fluctuations around the vacuum $\varphi=\varphi_{0}$
have to be taken into account, it follows from the constraint 
(\ref{IV26}) and Eq. (\ref{IV28}) that the nonminimal nonrenormalizable 
interaction of the gauge field with the scalar field appears.

\bigskip
\section{Model with a critical limit}

\bigskip   

The aim of this Appendix is to demonstrate reasons which
lead us to the model with persistent gauge condensate (Sec.IID).
In the framework of the strong NGVE principle we consider here the 
interesting family of Lagrangians  which 
depend on the ordinary gauge field $\tilde{A}_{\mu}$ and the three index 
gauge field $A_{\mu\nu\alpha}$ only through the gauge complex $y$, 
Eq.(\ref{V1}) with the power low dependence on $y$. Therefore, in the 
unitary gauge for $\tilde{A}_{\mu}$, instead of dealing with 
(\ref{IV23}),(\ref{IV24}), we have the action 
\begin{eqnarray}
S=\int\Phi d^{4}x\left[-\frac{1}{\kappa}R(\Gamma,g)
-\frac{1}{pm^{4(p-1)}}y^{p}+
\frac{1}{2}g^{\mu\nu}\varphi,_{\mu}\varphi,_{\nu}-
V(\varphi)+
\frac{1}{2}\tilde{e}^{2}\varphi^{2}g^{\mu\nu}\tilde{A}_{\mu}\tilde{A}_{\nu}
\right]
 \label{V2}
\end{eqnarray}
where dimensionless 
parameter $p$ is a real number. As we will see later, the physically 
interesting case is achieved in the critical limit $p\rightarrow\infty$.

Constraint (\ref{II3}) takes now the form
\begin{equation}
\frac{2p-1}{pm^{4(p-1)}}y^{p}=V(\varphi)+M
 \label{V3}
\end{equation}
which defines $y$ as a function of $\varphi$.

>From variation with respect to $A_{\nu\alpha\beta}$ we obtain the 
equation 
\begin{equation}
\partial_{\mu}(\chi y^{p-1}\varepsilon^{\mu\nu\alpha\beta})=0
 \label{V4}
\end{equation}
the solution of which can be written as
\begin{equation}
\chi y^{p-1}=\omega m^{4(p-1)}
 \label{V5}
\end{equation}
where $\omega$ is a dimensionless integration constant.

Varying with respect to the scalar field $\varphi$ we get
\begin{equation}
(-g)^{-1/2}\partial_{\mu}(\sqrt{-g}g^{\mu\nu}\partial_{\nu}\varphi)
+\sigma,_{\mu}\varphi ^{,\mu}+V^{\prime}(\varphi)+
\tilde{e}^{2}\varphi g^{\alpha\beta}\tilde{A}_{\alpha}\tilde{A}_{\beta}=0,
 \label{V6}
\end{equation}

The equation for the gauge field $\tilde{A}_{\alpha}$ is
\begin{equation}
\frac{1}{\sqrt{-g}}\partial_{\mu}[\chi y^{p-1}\sqrt{-g}
\tilde{F}^{\mu\nu}]
+\frac{\tilde{e}^{2}}{4}m^{4(p-1)}\varphi^{2}\chi\tilde{A}^{\nu}
=0
 \label{V7}
\end{equation}
which becomes
\begin{equation}
\frac{1}{\sqrt{-g}}\partial_{\mu}[\sqrt{-g}\tilde{F}^{\mu\nu}]
+\frac{\tilde{e}^{2}}{4\omega}\varphi^{2}\chi\tilde{A}^{\nu}
=0
 \label{V8}
\end{equation}
due to Eq. (\ref{V5}). And finally, the variation of $g^{\mu\nu}$ leads 
to the gravitational equations
\begin{eqnarray}
\frac{1}{\kappa}R_{\mu\nu}(\Gamma,g)=
-\frac{y^{p}}{2m^{4(p-1)}}g_{\mu\nu}+
\frac{y^{p-1}}{m^{4(p-1)}}\left[\frac{1}{2}\tilde{F}^{\alpha\beta}
\tilde{F}_{\alpha\beta}g_{\mu\nu}-
2\tilde{F}_{\mu\alpha}\tilde{F}_{\nu\beta}g^{\alpha\beta}\right]+
\frac{1}{2}\varphi_{,\mu}\varphi_{,\nu}+
\frac{\tilde{e}^{2}}{2}\varphi^{2}\tilde{A}_{\mu}\tilde{A}_{\nu}
 \label{V9}
\end{eqnarray}
where Eq. (\ref{V1}) has been used.

For the same reasons as those explained in Sec. IV, we have to perform 
the conformal transformation (\ref{IV11}) which provides a formulation of 
the theory in the Einstein picture.  Eqs. 
(\ref{V6}), (\ref{V8}) and (\ref{V9}) become then correspondingly
\begin{equation}
\frac{1}{\sqrt{-\overline{g}}}\partial_{\mu}(\sqrt{-\overline{g}}
\enspace\overline{g}^{\mu\nu}\partial_{\nu}\varphi)+
\frac{dV^{(p)}_{eff}}{d\varphi}+
e^{2}\varphi g^{\alpha\beta}A_{\alpha}A_{\beta}
=0
 \label{V10}
\end{equation}
\begin{equation}
\frac{1}{\sqrt{-\overline{g}}}\partial_{\mu}(\sqrt{-\overline{g}}
\enspace\overline{g}^{\mu\alpha}\overline{g}^{\nu\beta}
F_{\alpha\beta})+
e^{2}\varphi^{2}\overline{g}^{\nu\alpha}A_{\alpha}
=0
 \label{V11}
\end{equation} 
\begin{equation}
G_{\mu\nu}(\overline{g}_{\alpha\beta})=\frac{\kappa}{2}T_{\mu\nu}
 \label{V12}
\end{equation}
\begin{equation}
T_{\mu\nu}=
\varphi_{,\mu}\varphi_{,\nu}-
\frac{1}{2}g_{\mu\nu}\varphi_{,\alpha}\varphi_{,\beta}
\overline{g}^{\alpha\beta}+V^{(p)}_{eff}(\varphi)\overline{g}_{\mu\nu}+
\frac{1}{4}\overline{g}_{\mu\nu}F_{\alpha\beta}F^{\alpha\beta}
-F_{\mu\alpha}F_{\nu\beta}\overline{g}^{\alpha\beta}+
e^{2}\varphi^{2}(A_{\mu}A_{\nu}-
\frac{1}{2}\overline{g}_{\mu\nu}A_{\alpha}A^{\alpha})
 \label{V13}
\end{equation}
where
\begin{equation}
V^{(p)}_{eff}(\varphi)\equiv\left[\omega m^{4(1-1/p)}\right]^{-1}
\left(\frac{p}{2p-1}\right)^{2-1/p}(V(\varphi)+M)^{2-1/p}
 \label{V14}
\end{equation}
and the rescalings (\ref{Ch4})
have been performed. It is assumed that $\omega>0$. 

Equations (\ref{V10})-(\ref{V13}) describe the family of canonical 
equations of GR (parameterized by the parameter $p$) for a gauge model (in 
the unitary gauge) 
including gauge field $A_{\mu}$ minimally coupled to the scalar field
$\phi$ with the potential (\ref{V14}) and the coupling constant $e$. For 
$p=1$ (which has to be studied by itself), the theory reproduces the 
Einstein GR with the original potential $V(\varphi)$ and with a 
cosmological constant $M$ (see Ref. \cite{GK2}). In contrast, for any 
$p>1$, the effective potential $V^{(p)}_{eff}(\varphi)$, given by 
Eq.(\ref{V14}), has an absolute minimum at the point $\varphi_{0}$ where 
$V(\varphi_{0})+M=0$.

However an additional remarkable feature appears when 
$p\rightarrow\infty$. In this case
\begin{equation}
V_{eff}(\varphi)\equiv\lim_{p \to \infty}V^{(p)}_{eff}(\varphi)=
\frac{1}{4\omega m^{4}}(V+M)^{2}
 \label{V16}
\end{equation}
and for any analytical function $V(\varphi)$, all derivatives of the 
effective potential $V_{eff}(\varphi)$ 
are finite at the
absolute minimum $\varphi=\varphi_{0}$ where $V(\varphi_{0})+M=0$. In
particular, $V^{\prime\prime}_{eff}(\varphi_{0})\propto
[V^{\prime}(\varphi_{0})]^{2}$ is finite (and nonzero if we do not carry out
the fine tuning $V^{\prime}(\varphi_{0})=0$). Therefore the Higgs boson, in
particular, can reappear
as a physical particle of the theory. In the context of  cosmology
where $V_{eff}(\varphi)$ plays the role of the inflaton potential, a finite
mass of the inflaton allows to recover the usual oscillatory regime of
the reheating period after inflation that are usually considered.

Notice that again (as it was in Sec.IIID) the gauge coupling constant $e$ 
in the Einstein frame (see Eq. (\ref{Ch4})) depends on the integration 
constant $\omega$ which appears in the solution for the four index field 
condensate (\ref{V5}).

>From Eqs.(\ref{V3}) and (\ref{V5}) it follows that
\begin{equation}
\chi =(2-\frac{1}{p})^{1-1/p}\omega m^{4(1-1/p)}(V+M)^{-1+1/p}
 \label{V17}
\end{equation}
so that if $p>1$ or $p<0$ we obtain that $\chi\rightarrow\infty$ as 
$V+M\rightarrow 0$.

It is very instructive to look at what happens to the condensate $y$ when 
we approach the true vacuum $\varphi=\varphi_{0}$ where 
$V(\varphi_{0})+M=0$. For any finite $p>1$ we see from Eq. (\ref{V3}) 
that $y=(V+M)^{1/p}m^{4-4/p}(\frac{p}{2p-1})^{1/p}$ and $y\rightarrow 0$ 
in this limit. We notice however the very interesting effect which 
consists of the fact that as $p$ becomes big, $y$ approaches zero but at 
a very slow rate. In the limit $p\rightarrow\infty$ we can indeed argue 
that $y$ does not necessarily approaches zero but rather to an 
undetermined constant since $y\sim 0^{0}$ which is not defined. This 
suggests that the possible existence of a condensate $y$ that survives 
even in the true vacuum (which we will call "persistent condensate") is 
the cause of the remarkable feature which allows $V_{eff}$ to be of the 
form (\ref{V16}). In Sec. IID we have verified this 
explicitly.

\bigskip
\section{Constraint and LES in the Vierbein - Spin-Connection  
formalism}

\bigskip

To incorporate fermions into the NGVE theory we have to use the vierbein 
- spin-connection (VSC) formalism (see, however a purely affine approach 
due to Ne'eman \cite{Neeman}). In this Appendix we review in a short form 
how the NGVE principle works in the VSC-formalism.

 In this case we define \cite{Gasp} 
\begin{equation} 
R(\omega ,V) =V^{a\mu}V^{b\nu}R_{\mu\nu ab}(\omega),
\label{B1}
\end {equation}
\begin{equation}
R_{\mu\nu ab}(\omega)=\partial_{\mu}\omega_{\nu ab}
-\partial_{\nu}\omega_{\mu ab}+(\omega_{\mu a}^{c}\omega_{\nu cb}
-\omega_{\nu a}^{c}\omega_{\mu cb})
        \label{B2}
\end{equation}
where $V^{a\mu}=\eta^{ab}V_{b}^{\mu}$, $\eta^{ab}$ is the diagonal
$4\times 4$
matrix with elements $(1, -1,-1,-1)$ on the diagonal, $V_{a}^{\mu}$
are the vierbeins and $\omega_{\mu}^{ab}=-\omega_{\mu}^{ba}$ 
($a,b=0,1,2,3$) is
the spin connection. The matter Lagrangian $L_{m}$ that appears in
Eq.(\ref{Action}) is now a function of matter fields, vierbeins and spin
connection, considered as independent fields. We assume for simplicity
that $L_{m}$ does not depend on the derivatives of vierbeins and spin
connection.

We are now going to study the theory defined by the action 
(\ref{Action}) in the case that the scalar curvature is defined by 
(\ref{B1}),(\ref{B2}).

As in Sec. IIA, variation with respect to the scalar fields $\varphi_{a}$ 
leads to the equations
\begin{equation}
 A_{a}^{\mu}\partial_{\mu}(-\frac{1}{\kappa}R(V,\omega ) + L_{m}(V,\omega 
, matter fields)) = 0
 \label{B3}
 \end{equation}
which implies, if $\Phi \neq 0$, that

\begin{equation}
 -\frac{1}{\kappa}R(V,\omega ) + L_{m}(V,\omega, matter fields) = M
 \label{B4}
 \end{equation}

On the other hand, considering the equations obtained from the variation 
of the vierbeins, we get if  $\Phi \neq 0$
\begin{equation}
 -\frac{2}{\kappa}R_{\mu a}(V,\omega)
+\frac{\partial L_{m}}{\partial V^{a\mu}} = 0,
 \label{B5}
 \end{equation}
where
\begin{equation}
 R_{\mu a}(V,\omega)\equiv 
V^{b\nu}R_{\mu\nu ab}(\omega).
\label{B6}
\end{equation}

Notice that eq.(\ref{B5}) is indeed invariant under the shift 
$L_{m}\rightarrow L_{m}+const$. By using Eq. (\ref{B1}) we can eliminate 
$R(\omega)$ from the equations (\ref{B4}) and (\ref{B5}) after contracting 
the last one with $V_{a\mu}$. As a result we obtain {\em the nontrivial 
constraint} in the form 

\begin{equation}
V^{a\mu}\frac{\partial(L_{m}-M)}{\partial V^{a\mu}}-2(L_{m}-M)=0
\label{B7}
 \end{equation}
which replaces Eq. (\ref{II3}) (see Sec. II) and in the absence of 
fermions, is equivalent to the constraint (\ref{II3}).

The simplest example of a fermion is that of spin 1/2 particles. In
this case we regard the spinor field $\Psi$  as a general
coordinate
scalar and transforming nontrivially with respect to local Lorentz
transformation according to the spin $1/2$ representation of the Lorentz
group.

The NGVE principle prescripts for the fermionic  hermitian action (which 
allows for the possibility of fermion self interactions) to be of the form
\begin{equation}
S_{f}=\int L_{f}\Phi d^{4}x
 \label{B8}
 \end{equation}
where
\begin{equation}
 L_{f}=\frac{i}{2}\overline{\Psi}
[\gamma^{a}V_{a}^{\mu}(\vec\partial
_{\mu}+\frac{1}{2}\omega_{\mu}^{cd}\sigma_{cd})
-(\overleftarrow{\partial}_{\mu}-\frac{1}{2}\omega_{\mu}^{cd}\sigma_{cd})
\gamma^{a}V_{a}^{\mu}]\Psi+U(\overline{\Psi}\Psi)
 \label{B9}
 \end{equation}
Here $\sigma_{cd}\equiv \frac{1}{4}[\gamma_{c},\gamma_{d}]$.
Spin-connection $\omega_{\mu}^{cd}$ should be determined by the equation 
obtained
from the variation of the full action with respect to $\omega_{\mu}^{cd}$ 
(see Appendix F). 

>From (\ref{B9}) and using the equations of motion 
derived from the action (\ref{B8}),(\ref{B9}), we get
\begin{equation}
V_{a}^{\mu}\frac{\partial L_{f}}{\partial V_{a}^{\mu}}-2L_{f}=
\overline{\Psi}\Psi U'-2U,
 \label{B10}
 \end{equation}
where $U'$ is the derivative of $U$ with respect to its argument 
$\overline{\Psi}\Psi$. We see
that the constraint (\ref{B7}) is satisfied on the mass shell (since the
fermion equations of motion are used) with $M=0$ for $L_{f}$
defined by eq.(\ref{B9}) if, for example, $U=c(\overline{\Psi}\Psi)^{2}$.
Any other quartic interaction, like
$\overline{\Psi}\gamma_{a}\Psi\overline{\Psi}\gamma^{a}\Psi$,
$\overline{\Psi}\sigma_{ab}\Psi\overline{\Psi}\sigma^{ab}\Psi$,
$(\overline{\Psi}\gamma_{5}\Psi)^{2}$, etc. would also satisfy the
constraint (\ref{B7}) on the mass shell with $M=0$. In particular, the
Nambu - Jona-Lasinio
model\cite{NJL} would also satisfy the constraint (\ref{B7}) on the mass
shell with $M=0$.

In the presence of Dirac fermions  with the Lagrangian density (\ref{B9}),
the LES  (\ref{ES1}), (\ref{ES2}) is appropriately
generalized to
\begin{equation}
V^{a}_{\mu}(x)=J^{-1/2}(x)V^{\prime a}_{\mu}(x);\qquad V_{a}^{\mu}(x)=
J^{1/2}(x)V^{\prime\mu}_{a}(x)
\label{B11}
\end{equation}
\begin{equation}
\Phi(x)=J^{-1}(x)\Phi^{\prime}(x)
\label{B12}
\end{equation}
\begin{equation}
\psi(x)=J^{1/4}(x)\psi^{\prime}(x);\qquad
\overline{\psi}(x)=J^{1/4}(x)\overline{\psi}^{\prime}(x)
\label{B13}
\end{equation}
provided that $V(\overline{\psi}\psi)$ has a form of one of the mentioned
above quartic interactions.
Notice that in this case the condition for the invariance of the action with
the matter Lagrangian
(\ref{B9}) under the transformations (\ref{B11})-(\ref{B13}) is not just
the simple homogeneity of degree 1 in $g^{\mu\nu}$ or degree 2 in
$V_{a}^{\mu}$, because of the presence of the fermion transformation
(\ref{B13}).

\bigskip
\section{Connection in the VSC-formalism}
\bigskip

We analyze here what is the dependence of the spin connection 
$\omega_{\mu}^{ab}$ on $V^{a}_{\mu}$, $\chi$, $\Psi$ and $\overline{\Psi}$.
Varying the action (\ref{VI1}) with respect to 
$\omega_{\mu}^{ab}$ and making use that
\begin{equation}
R(V,\omega)\equiv 
-\frac{1}{4\sqrt{-g}}\varepsilon^{\mu\nu\alpha\beta}\varepsilon_{abcd}
V^{c}_{\alpha}V^{d}_{\beta}R_{\mu\nu}^{ab}(\omega)
\label{C1}
 \end{equation}
we obtain
\begin{equation}
\varepsilon^{\mu\nu\alpha\beta}\varepsilon_{abcd}[\chi 
V^{c}_{\alpha}D_{\nu}V^{d}_{\beta}
+\frac{1}{2}V^{c}_{\alpha}v^{d}_{\beta}\chi,_{\nu}]+
\frac{\kappa}{4}\sqrt{-g}V^{c\mu}\varepsilon_{abcd}\overline{\Psi}
\gamma^{5}\gamma^{d}\Psi=0, 
\label{C2}
 \end{equation}  
where 
\begin{equation}
D_{\nu}V_{a\beta}\equiv\partial_{\nu}V_{a\beta}
+\omega_{\nu a}^{d}V_{d\beta}
\label{C3}
 \end{equation}
The solution of Eq. (\ref{C2}) is represented in the form
\begin{equation}
\omega_{\mu}^{ab}=\omega_{\mu}^{ab}(V) + K_{\mu}^{ab}(\sigma)  + 
K_{\mu}^{ab}(V,\overline{\Psi},\Psi)
\label{C4}
 \end{equation}
where 
\begin{equation}
\omega_{\mu}^{ab}(V)=V_{\alpha}^{a}V^{b\nu}\{ ^{\alpha}_{\mu\nu}\}-
V^{b\nu}\partial_{\mu}V_{\nu}^{a}
\label{C5}
 \end{equation}
is the Riemannian part of the connection,
\begin{equation}
K_{\mu}^{ab}(\sigma)=\frac{1}{2}\sigma_{,\alpha}(V_{\mu}^{a}V^{b\alpha}-
V_{\mu}^{b}V^{a\alpha})
\label{C6}
 \end{equation}
and
\begin{equation}
K_{\mu}^{ab}(V,\overline{\Psi},\Psi)=
\frac{\kappa}{8}\eta_{ci}V_{d\mu}\varepsilon^{abcd}\overline{\Psi}
\gamma^{5}\gamma^{i}\Psi.
\label{C7}
 \end{equation}
Notice that the spin-connection $\omega_{\mu}^{ab}$ defined by Eqs. 
(\ref{C4})-(\ref{C7}) is invariant under the LES transformations 
(\ref{B11})-(\ref{B13}).


\bigskip
\begin{thebibliography}{99}
\bigskip

\bibitem{CC}
I. Novikov, {\it Evolution of the Universe,}
Cambridge University Press, 1983.
S.Weinberg, Rev. Mod. Phys. {\bf 61}, 1 (1989); Y.J. Ng, Int. J. Mod.
Phys. {\bf D1}, 145, (1992); {\it Gravitation and Modern Cosmology, The
Cosmological Constant Problem}, edited by A.Zichichi, V. deSabbata and N.
Sanchez (Ettore Majorana International Science Series, Plenum Press, 1991).
M. Veltman, "Reflections on the Higgs system". Preprint CERN 97-05.

\bibitem{GK1}
E.I. Guendelman and A.B. Kaganovich, Phys. Rev. {\bf D53}, 7020
(1996). See also in:{\it
Proceedings of
the third Alexander Friedmann International Seminar on Gravitation and
Cosmology}, edited by Yu.N.Gnedin, A.A. Grib, V.M. Mostepanenko (Friedman
Laboratory Publishing, St Petersburg, 1995).

\bibitem{GK2}
E.I. Guendelman and A.B. Kaganovich,  Phys. Rev.{\bf D55}, 5970 (1997);
E.I. Guendelman and A.B. Kaganovich, Mod. Phys. Lett. {\bf A12}, 2421
(1997).

\bibitem{GK3}
E.I. Guendelman and A.B. Kaganovich, Phys. Rev.{\bf D56}, 3548 (1997).
E.I. Guendelman and A.B. Kaganovich, Hadronic Journal {\bf 21}, 19 (1998).

\bibitem{GK4}
E.I. Guendelman and A.B. Kaganovich, Mod. Phys. Lett. {\bf A13}, 1583
(1998).

\bibitem{Hehl}
F. Gronwald, U. Muench and F. W. Hehl, Hadronic Journal, {\bf 21}, 3 (1998).

\bibitem{GK5}
E.I. Guendelman and A.B. Kaganovich, Phys. Rev. {\bf D57}, 7200 (1998).

\bibitem{PASCOS}
E.I. Guendelman and A.B. Kaganovich, "Gravity, Cosmology and Particle
Fields Dynamics without the Cosmological Constant Problem", to appear in
the Proceedings of the sixth International Symposium on Particles,
Strings and Cosmology, PASCOS-98. 


\bibitem{Fried98}
E.I. Guendelman and A.B. Kaganovich, "Field Theory Models  without the
Cosmological Constant Problem", Plenary talk (given by E. I. Guendelman) 
on the fourth Alexander
Friedmann International Seminar on Gravitation and
Cosmology; gr-qc/9809052.

\bibitem{HN}
F. W. Hehl, J. D. McCrea, E. W. Mielke, Yuval Ne'eman,
Phys. Reports, {\bf 258},1 (1995).

\bibitem{EK}
A.Einstein, {\it The Meaning of Relativity}, Fifth Edition, MJF
Books, N.Y.1956   ( see Appendix II).

\bibitem{Gut}
A. H. Guth, Phys. Rev. {\bf D23}, 347 (1981); A. D. Linde, Phys. Lett.
{\bf 108B}, 389 (1982); A. Albrecht and P. J. Steinhardt, Phys. Rev.
Lett. {\bf 48}, 1220 (1982), A. A. Starobinsky, Phys. Lett. {\bf B91}, 99
(1980); A. D. Linde, Phys. Lett. {\bf B129}, 177 (1983); D. La and D. J.
Steinhardt, Phys. Rev. Lett. {\bf 62}, 376 (1989).


\bibitem{KT}
See, for example: E. Kolb and M. S. Turner,{\it The Early Universe}, Addison
Wesley, 1990 (see p.314).

\bibitem{G1}
E.I. Guendelman, "Scale invariance, new inflation and decaying Lambda
terms", gr-qc/9901017, to appear in Mod. Phys. Lett. A.

\bibitem{G11}
E.I. Guendelman,
 "Scale invariance, mass and cosmology",
gr-qc/9901067.

\bibitem{G}
E.I. Guendelman, Phys. Lett. {\bf B412}, 42 (1997)

\bibitem{Gasp}
See for example, P.G.O.Freund, {\it Introduction to Supersymmetry},
Cambridge
University Press, 1986, Chapt.21; V.de Sabbata and M.Gasperini,
{\it Introduction to Gravitation}, World Scientific (1985); B.de Wit and
D.Z.Freedman, "Supergravity-The basics and beyond", MIT preprint
CPT N1238, January 1985.

\bibitem{NJL}
Y.Nambu and G.Jona-Lasinio, Phys. Rev. {\bf122}, 345 (1961).

\bibitem{Nambu-Bardeen}
Y. Nambu, Proceedings of the 1988 Kazimierz Workshop, Z. Ajduk et al., 
eds. (World Scientific, 1989); Proceedings of the 1988 Nagoya Workshop, 
M. Bando et al., eds. (World Scientific, 1989). W.Bardeen, C. Hill and 
M.Linder, Phys. Rev. {\bf D41}, 1647 (1990)

\bibitem{Oku}
L. B. Okun, {\it Leptons and Quarks}, Amsterdam, North Holland, 1982.

\bibitem {Col}
S. Coleman, {\em Aspects of Symmetry}, Cambridge Univ. Press,
Cambridge,1985.

\bibitem{worm}
S. Coleman, Nucl. Phys. {\bf B310}, 643 (1988); J. Preskill, ibid. {\bf
B323}, 141 (1989); G. B. Giddings and A. Strominger, ibid., {\bf B307},
854 (1988); S. W. Hawking, ibid., {\bf B335}, 155 (1990).

\bibitem{QCDcondens}
The early  works on this subject are:
G. K. Savvidy, Phys. Lett. {\bf B71}, 133 (1977); N. K. Nielsen and P.
Olesen, Nucl. Phys. {\bf B144}, 376 (1978); N. K. Nielsen and P.
Olesen, Phys. Lett. {\bf B79}, 304 (1978).

\bibitem{GNP}
E. I. Guendelman, E. Nissimov, S. Pacheva, Phys. Lett., {\bf B360}, 57
(1995).

\bibitem{Castro}
C. Castro, Int. Journ. of Mod. Phys., {\bf A13}, 1263 (1998).

\bibitem{G2}
E. I. Guendelman, "Gauge condensates and gauge dynamics, the cosmological
and strong CP problem", to appear in Int. J. Mod. Phys.A.

\bibitem{nonmetr}
J. A. Schouten, {\it Ricci-Calculus}, Springer-Verlag, Berlin, 1954.
P. Baekler, F. W. Hehl and E. W. Mielke, "Nonmetricity and torsion: Facts
and Fancies in gauge approaches to gravity", in: {\it Procc. of the Fourth
Marcel Grossman Meeting on General Relativity}, ed. by R. Rufini,
Elsevier Sc., 1986.

\bibitem{FF1}
A. Aurilia, H. Nicolai and P. K. Townsend, Nucl.Phys.
{\bf B176}, 509 (1980); S. W. Hawking, Phys. Lett., {\bf 134B}, 276
(1984);
E. Witten , in Shelter II 1985 {\it Proc. 1983 Shelter
Island
Conference on Quantum Field Theory and the Fundamental Problems of
Physics} (Cambridge, MA: MIT Press).

\bibitem{FF2}
A. Aurilia, D. Christodoulou and F. Legovini, Phys. Lett. {\bf 73B}, 429
(1978); A. Aurilia, G. Denardo, F. Legovini and E. Spalluci, ibid., {\bf
147B}, 258 (1984);
 J. D. Brown and C. Teitelboim,
Phys. Lett. {\bf 195B}, 177 (1987); Nucl. Phys. {\bf B297}, 787 (1988).

\bibitem{Neeman}
 Y. Ne'eman, Ann. Inst. Henri Poincare {\bf XXVIII}, 369 (1978); Y.
Ne'eman and Dj.Sijacki, Ann. Phys. {\bf 129}, 292 (1979).


\end{thebibliography}
\end{document}